\documentclass[11pt%
]{article}
\usepackage{cite}
\usepackage{amsmath}
\usepackage{amsfonts}
 \usepackage{slashed}
\usepackage{amssymb}
\usepackage{graphicx}
\usepackage{dcolumn}
\usepackage{bm}
\usepackage{hyperref}
\usepackage{pstricks}
\usepackage{color}
\usepackage{cancel}

\usepackage{amsmath}
\usepackage{amsfonts}
\usepackage{slashed}
\usepackage{amssymb}
\usepackage{wasysym}
\usepackage{graphicx}
\usepackage{dcolumn}
\usepackage{bm}
\usepackage{hyperref}
\usepackage{pstricks}
\usepackage{cancel}
\usepackage{color,xcolor,soul}

\newcommand{\lsim}{\;\raisebox{-.3em}{$\stackrel{\displaystyle <}{\sim}$}\;}
\newcommand{\gsim}{\;\raisebox{-.3em}{$\stackrel{\displaystyle >}{\sim}$}\;}
\newcommand{\mev}{~\rm MeV}
\newcommand{\gev}{~\rm GeV}
\newcommand{\tev}{~\rm TeV}

\newcommand{\mdm}{m_{H^0}}
\newcommand{\mhp}{m_{H^\pm}}
\newcommand{\mha}{m_{A^0}}
\newcommand{\msig}[2]{m_{\Sigma_{#1}^{#2}}}

\newcommand{\ZZ}{$ \mathbf{Z}_2$}

\newcommand{\CheckMATE}{\tt CheckMATE}

\newcommand{\SW}{\mathcal{S}_W}
\newcommand{\SB}{\mathcal{S}_B}
\evensidemargin \oddsidemargin
\allowdisplaybreaks

\begin{document}

\date{\today}%

\begin{center}
{\large\sc {\bf Radiative Type III Seesaw Model and its collider phenomenology}
\vspace*{.4cm} }

{\sc
{Federico von der Pahlen}%
$^{a,}$\footnote{email: federico.vonderpahlen@udea.edu.co},~%
{Guillermo Palacio}%
$^{a,}$\footnote{email: galberto.palacio@udea.edu.co},~%
{Diego Restrepo}%
$^{a,b,}$\footnote{email: restrepo@udea.edu.co}~,%
\\
and {Oscar Zapata}%
$^{a,}$\footnote{email: oalberto.zapata@udea.edu.co}%
}
\vspace*{.4cm}
 
{\sl $^{a}$ Instituto de F\'isica, Universidad de Antioquia, Calle 70 No.\ 52-21, Medell\'in, Colombia}
\\
\vspace*{.4cm}
{\sl $^{b}$ Simons Associate at ICTP The Abdus Salam International Centre for Theoretical Physics, \\
Strada Costiera 11, I-34151, Trieste, Italy}

\end{center}

\begin{abstract}
We analyze the present bounds of a scotogenic model, the Radiative Type III Seesaw (RSIII), 
in which an additional scalar doublet and at least two fermion triplets of $SU(2)_L$ 
are added to the Standard Model (SM).
In the RSIII the new physics (NP) sector is odd under an exact global \ZZ\ symmetry.
This symmetry guaranties that the lightest NP neutral particle is stable, providing a natural dark matter (DM) candidate, 
and leads to naturally suppressed neutrino masses generated 
by a one-loop realization of an effective Weinberg operator.
We focus on the region with the highest sensitivity in present and future LHC searches, 
with light scalar DM and at least one NP fermion triplet at the sub-TeV scale.
This region allows for significant production cross-sections of NP fermion pairs at the LHC.
We reinterpret a set of searches for supersymmetric particles at the LHC obtained using the package {\CheckMATE}, 
to set limits on our model as a function of the masses of the NP particles and their Yukawa interactions.
The most sensitive search channel is found to be dileptons plus missing transverse energy.
In order to target the case of tau enhanced decays and the case of compressed spectra 
we reinterpret the recent slepton and chargino search bounds by ATLAS.
For a lightest NP fermion triplet with a maximal branching ratio to either electrons or muons
we exclude NP fermion masses of up to $650\gev$, while this bound is reduced to approximately $400\gev$ in the tau-philic case.
Allowing for a general flavor structure we set limits on the Yukawa couplings, 
which are directly related to the neutrino flavor structure.
\end{abstract} 

\setcounter{footnote}{0}
\newpage
\section{\label{sec:intro}Introduction}

One of the simplest models which explains the dark matter (DM) content of the Universe is the Inert Doublet Model (IDM)%
~\cite{Deshpande:1977rw,Barbieri:2006dq},
where an additional scalar doublet of $SU(2)_L$ odd under a global \ZZ\ is added to the Standard Model (SM). 
The observed relic density of DM~\cite{Ade:2015xua} can be obtained in two regions of parameter space%
~\cite{LopezHonorez:2006gr,Dolle:2009fn,Honorez:2010re,LopezHonorez:2010tb,Sokolowska:2011sb,Gustafsson:2012aj,Goudelis:2013uca},
the low mass region, for DM masses around the Higgs resonance, and the high mass region, for DM masses above $500\gev$. 
In the former, the reach at the LHC is quite restricted by the large backgrounds coming from gauge final states~\cite{Belanger:2015kga,Poulose:2016lvz}, 
while in the latter, the reach is limited by the small cross sections and small mass splittings required to explain the observed DM relic density~\cite{Queiroz:2015utg}. 
The IDM is also in agreement with direct detection limits%
~\cite{Arhrib:2013ela,Klasen:2013btp,Abe:2015pra,Abe:2014gua,Abe:2015rja,Ilnicka:2015jba},
limits on indirect detection in gamma rays~\cite{Gustafsson:2007pc,Ilnicka:2015jba},
limits on indirect detection in neutrino telescopes~\cite{Agrawal:2008xz},
LEP searches~\cite{Lundstrom:2008ai},
and dilepton searches at the LHC~\cite{Dolle:2009ft,Belanger:2015kga}.
The introduction of at least two additional \ZZ-odd fermion singlets 
opens the possibility to explain the smallness of neutrino masses 
through radiative corrections at one-loop in the IDM~\cite{Ma:2006km}.
The same symmetry which guarantees DM stability also forbids the tree level contribution to neutrino masses.  
In this way, the so-called scotogenic model constitutes a 
solid framework to explain simultaneously DM and radiative neutrino masses.  
The minimal scotogenic model with singlet fermions is not the only possibility 
to explain both radiative neutrino masses and the correct DM relic density%
~\cite{Ma:2008cu,Hambye:2009pw,Gustafsson:2012vj,Chao:2012sz,Vicente:2014wga,Merle:2015gea,Okada:2014qsa,Longas:2015sxk,Chakrabarty:2015yia}.
The realization of the Weinberg operator at one-loop involves the generic coupling of the lepton doublets 
with both \ZZ-odd scalar and fermion multiplets~\cite{Restrepo:2013aga}. 
Including larger fermion representations also implies that these fermions interact with the electroweak gauge bosons,
leading to significantly large production cross-sections at the LHC.
This is in stark contrast with the minimal scotogenic model, 
where the singlet fermions cannot be directly produced, 
resulting in a very limited collider phenomenology%
\footnote{It should be noticed that 
significant production of the fermion singlets can be achieved 
in some regions of the parameter space with very light DM where the full relic density cannot be accounted for~{\cite{Sierra:2008wj}}.}.
For suitable choices of the spectrum and sufficiently high $SU(2)_L$ representations
these new Yukawa interactions lead to the decay of the \ZZ-odd fermions, 
opening the possibility to generate collider signals of dileptons plus missing transverse energy (MET).

Along this idea, the simplest extension of the minimal scotogenic model consists in 
replacing at least one of the fermion singlets by a fermion triplet~\cite{Ma:2008cu,Law:2013saa,Hirsch:2013ola}.
This model leads to the same neutrino masses and DM relic density
but has a richer collider phenomenology with strong similarities with the minimal supersymmetric standard model (MSSM)%
~\cite{Nilles:1983ge,Haber:1984rc,Barbieri:1987xf}.
Both in the extended scotogenic model as in the MSSM the fermion triplets may be produced in pairs 
in Drell-Yan processes, resulting in production cross-sections of the same order.
At the LHC the two cross-sections are equal in the well-studied wino limit with decoupled Higgsinos and squarks,
where the lightest chargino and the second lightest neutralino are wino-like.
In this limit the t and u-channel amplitudes in the MSSM processes can be neglected
and the production of the \ZZ-odd fermions proceeds via gauge boson exchange in s-channel.
The leading one-loop QCD corrections for final fermion states with the same $SU(2)_L$ quantum numbers
are also the same since only the initial quarks are involved.
The analogy with the MSSM also implies that, in the case that the fermion triplet is the lightest \ZZ-odd state,
requiring sufficiently high relic density abundance forces these fermions to be heavy~\cite{Chao:2012sz}, 
above the mass reach of the LHC. 
We focus instead on a simplified model scenario with scalar DM 
in which the neutral component of the fermion triplet is the next to lightest \ZZ-odd particle (NLOP).
The charged triplet components are slightly heavier since the degeneracy is broken at one-loop level.
In this simplified model scenario DM limits only constrain the scalar sector of the model,
allowing for scenarios where the \ZZ-odd fermions may be light enough to be copiously produced at the LHC.
This motivates analyses designed to constrain these models in present LHC searches and to potentially determine their existence in the future.

The decays of both scotogenic and supersymmetric particles are constrained by the \ZZ\ symmetry, 
which leads to cascades to the lightest odd particle (LOP), with the resulting MET signature. 
In the framework of simplified model searches at the LHC limits for sleptons and electroweakinos in the MSSM have been given 
for different spectra, characterized by
 sleptons being either lighter or heavier than the wino-like charginos and neutralinos~%
\cite{ATLAS-CONF-2013-049,Aad:2014nua,Aad:2014vma,Khachatryan:2014qwa,Aad:2015eda}. 
The strong similarity with our simplified model allows to reinterpret those limits for processes with the same decay topologies.
In the scotogenic simplified model defined by NLOP \ZZ-odd fermions triplets, the latter
decay to the DM candidate and a lepton.
The collider signature at the LHC from 
charged \ZZ-odd fermion pair-production is opposite sign dileptons plus MET.
The flavor of the decay leptons is determined by the NP Yukawa couplings, 
which are in turn related to the neutrino mass generating operators constrained by neutrino experiments~\cite{Forero:2014bxa}.
A determination of the flavor structure of the final state 
is therefore highly relevant, and may additionally allow to distinguish between different models.

Several supersymmetric processes lead to similar collider signatures at the LHC as the simplified scotogenic model, 
albeit with a different flavor structure.
Production of a chargino-neutralino pair decaying to intermediate sleptons leads to the so-called trilepton ``golden channel'', 
with the highest exclusion sensitivity in electroweakino searches. 
If one of the final leptons is lost this process may lead to opposite sign different flavor (OSDF) 
or opposite sign same flavor (OSSF) leptons plus MET.
Chargino pairs decaying to a lepton and a slepton, or sleptons pairs decaying to the neutralino and a lepton
are optimized in LHC searches for signal regions (SRs) with opposite sign same flavor (OSSF) leptons plus MET%
\footnote{As in most analyses we only consider the case of the minimal flavor violating MSSM where the slepton mass matrices are flavor diagonal.}. 
The decay topology of the sleptons 
is the same as that of the fermions of the scotogenic model.
The former have smaller production cross-sections, mainly due to the smaller number of spin degrees of freedom of the scalars. 
ATLAS~\cite{ATLAS-CONF-2013-049,Aad:2014vma} and CMS~\cite{Khachatryan:2014qwa} analyses 
searching for left-handed sleptons of the first two families with light neutralinos constrain masses below roughly $300\gev$.
Assuming that the detection efficiency of the most sensitive signal region (SR) in these analyses remains constant up to higher mass scales
one can estimate a lower mass exclusion limit  of around $630\gev$
for the \ZZ-odd triplet fermions decaying to only one lepton flavor.
A more precise limit can be obtained reinterpreting the recent SUSY searches in the framework of simplified models
with help of some of the recent high energy physics tools. 
The package {\CheckMATE}~\cite{Drees:2013wra,Cacciari:2005hq,Cacciari:2008gp}
allows to obtain exclusion limits
on supersymmetric simplified models based on an increasing number of ATLAS and CMS analyses.
It also allows to implement new physics models, resulting in exclusion limits based on the collider signatures of the
experimental analyses. 
It is therefore a useful approach in scotogenic models since
similar production and decay topologies as in the MSSM lead to similar collider signatures.
In particular, one may analyze the exclusion sensitivity as a function of the flavor space, 
which is determined by new Yukawa couplings between the \ZZ-odd fields and the leptons.
Decays with taus in the final state have a much lower exclusion sensitivity. 
Presently only upper limits on stau production cross-sections
 have been reported by dedicated analyses for stau production by ATLAS~\cite{Aad:2015eda} and CMS~\cite{CMS:2016saj}.
However, taking into account the larger cross-section for fermion pairs 
and recasting those results accordingly may allow to exclude light fermions 
decaying exclusively into taus and MET above the LEP exclusion limit~\cite{LEPSUSYWG} 
up to a lower mass limit of roughly $400\gev$. 
One can thus set solid exclusion bounds within the simplified scotogenic model
and  full flavor space allowed by neutrino physics
since final states with taus have the lowest exclusion sensitivity.

More complex decay chains open up when several scalars are lighter than the decaying 
\ZZ-odd fermions. In this case, the simplest exclusion limits can be obtained 
considering only those fermion decays to the DM candidate and a lepton.
This is equivalent to rescaling the production cross-section with the decay branching ratio~\cite{Bharucha:2013epa}, 
with the resulting loss in exclusion sensitivity.
The least convenient scenario in this respect corresponds to nearly degenerate scalars, with the neutral scalar mass splitting large enough
to generate additional hadronic activity. In this case the useful fermion branching ratio is reduced by a factor of almost four.
It may therefore be possible to set limits on this last scenario, and thus to all intermediate cases.
Another possible scenario is that when more than one fermion triplet is produced,
analogous to the supersymmetric case
when more than one slepton pair is kinematically available, where more stringent limits may be possible than with only one family.

This work is organized as follows. 
In Sec.~\ref{sec:Model} we introduce our model. 
In Sec.~\ref{sec:Model_Constraints} we analyze the constraints on the model and their implications for the low DM mass region.
In Sec.~\ref{sec:pheno} we discuss its collider phenomenology and our strategy to set limits on the model.
In Sect.~\ref{sec:results} we discuss our numerical results
and finally we summarize in Sect.~\ref{sec:level5}.

\section{The Model }\label{sec:Model} 

In this section we introduce the model RSIII~\cite{Ma:2008cu,Chao:2012sz,Restrepo:2013aga}, 
an extension of the SM with an additional complex scalar doublet of $SU(2)$, $\Phi$, 
and $n_{\Sigma}\ge 2$ generations of vector fermion triplets of $SU(2)$, 
$\Sigma_{k}$, $k=1,\ldots,n_{\Sigma}$.
The quantum numbers of the scalar and leptonic sector of the model
are given in Table~\ref{tab:particles}.
The new particles are odd under an exact \ZZ~symmetry,
forcing the lightest \ZZ-odd particle to be stable,
and thus a natural DM candidate.
This symmetry also prevents neutrino masses from being generated by the 
tree-level Type III seesaw mechanism~\cite{Foot:1988aq}, 
only allowing for the one-loop realization of the Weinberg operator.
Neutrino masses are generated at the one-loop level~\cite{Ma:2006km} 
via their interactions with the neutral components of $\Sigma_{k}$, 
the Majorana fermions $\Sigma_{k}^{0}$, 
and the neutral components of $\Phi$, $\phi^{0}$.
Therefore, the \ZZ~symmetry plays a crucial role linking DM to the neutrino mass generation%
\footnote{It is worth mentioning that the evolution of 
the model parameters via the renormalization group equations may induce a non-zero 
vacuum expectation value for $\phi^{0}$ at high scales, 
leading to the spontaneous breaking of the \ZZ~symmetry.
This situation, that indeed occurs in the minimal scotogenic model \cite{Merle:2015gea}, 
may be naturally avoided 
extending the model with a \ZZ-even real scalar-triplet,
as shown in \cite{Merle:2016scw} in the 
context of the scotogenic model 
where a fermion singlet is replaced by a fermion triplet~\cite{Hirsch:2013ola}.
This solution, where the evolution of
 the couplings of the scalar sector is modified by the extention of the scalar sector,
is fully applicable to our case.}.

\subsection{Lagrangian}
The most general renormalizable Lagrangian of the RSIII reads 
\begin{eqnarray}\label{eq:LRSIII}
\mathcal{L_{\rm RSIII}} = \mathcal{L}_{\rm{SM}} + \mathcal{L}_{\rm{NP}} 
~,
\end{eqnarray}
with~\cite{Chao:2012sz}
\begin{align}
\label{eq:LNP}
\mathcal{L}_{\rm{NP}} = &\ 
   i {\rm Tr}\left[\overline{\Sigma} \slashed D\Sigma\right] 
   - \dfrac{1}{2}{\rm Tr}\left[
     \overline{\Sigma} M_{\Sigma} \Sigma^{c} + \overline{\Sigma^{c}} M_{\Sigma}^{*} \Sigma
                        \right] 
   - \left(      
     Y_{k\alpha}^{} 
    \widetilde{{\Phi}}^{\dagger}\overline{{\Sigma}_k}{L}_{\alpha}
     + {\rm h.c.}
    \right) 
\nonumber\\ & + (D_\mu\Phi)^\dagger(D^\mu\Phi)- V_{\rm NP}({\Phi},\Phi_{\rm SM}) 
\,
,    
\end{align}  
with $\alpha=e,\mu,\tau$. 
Here the trace runs over the $SU(2)$ indices, 
the mass matrix $M_{\Sigma}$ (but not the NP Yukawa couplings $Y$)
is assumed to be
flavor diagonal, 
$ D $ denotes the covariant derivative, 
$L$ are the left-handed lepton doublets,
and  $\Phi_{\rm SM}$ is the SM scalar doublet.
Whenever possible the flavor indices have been suppressed. 
The NP scalar potential is given by
\begin{align}
\label{eq:VNP}
 V_{\rm NP}({\Phi},\Phi_{\rm SM}) 
 =  & \ \mu_{2}^{2}{\Phi}^{\dagger}{\Phi} 
 + 
  \lambda_{2}({\Phi}^{\dagger}{\Phi})^{2} 
 + {\lambda_{3}}(\Phi_{\rm SM}^{\dagger}\Phi_{\rm SM})({\Phi}^{\dagger}{\Phi}) 
\nonumber \\ &
 + {\lambda_4}(\Phi_{\rm SM}^{\dagger}{\Phi})({\Phi}^{\dagger}\Phi_{\rm SM}) 
+ \dfrac{\lambda_5}{2} \left[
(\Phi_{\rm SM}^{\dagger}{\Phi})^{2} +{\rm h.c.}\right]
\, ,\quad  
\end{align}  
with all the scalar couplings $\lambda_{i}$ real.

The  scalar fields are given by
\begin{eqnarray}\label{eq:Phi.expand}
{\Phi} = \begin{pmatrix}
H^{+} \\ 
\dfrac{1}{\sqrt{2}} ({H^{0} + iA^{0} })
\end{pmatrix},
\hspace{1cm}
\Phi_{\rm SM} = \begin{pmatrix}
G^{+} \\ 
\dfrac{1}{\sqrt{2}} ({v+ h + iG_{I}^{0} })
\end{pmatrix}, 
\end{eqnarray}
where $G_{I}^{0}$ and $G^{+} $ the Goldstone bosons of the SM, 
$ \langle \Phi_{\rm SM} \rangle  =\begin{pmatrix}
0 , & {v}/{\sqrt{2}}
\end{pmatrix}^{T}  $ 
with $v = 246\gev$. 
The masses for the NP scalars can be obtained from Eq.~(\ref{eq:VNP}):
\begin{eqnarray}\label{eq:Mscalars}
 \nonumber
 \ \mhp^{2} & = &  \mu_{2}^{2} +\dfrac{\lambda_{3}}{2} v^{2} ~,
\\ \nonumber
 \ \mdm^{2} & = & \mu_{2}^{2} +\dfrac{( \lambda_{3} + \lambda_{4} + \lambda_{5})}{2} v^{2} ~,
\\ 
 \ \mha^{2}  & = &  \mu_{2}^{2} +\dfrac{( \lambda_{3} + \lambda_{4} - \lambda_{5})}{2} v^{2} 
~. 
\end{eqnarray}  
where
$H^{0}$ and $A^{0}$ denote the neutral scalar and pseudoscalar components of the \ZZ-odd  scalar, 
and $H^{\pm}$ its charged components.
The Higgs mass is fixed to its current experimental value measured by ATLAS and CMS,
$m_h=125.09\pm 0.24\gev$~\cite{Aad:2015zhl}.
\begin{table}
\caption{Gauge, \ZZ\ and spin quantum numbers of the particle content of the RSIII entering $\mathcal{L}_{\rm{NP}}$,  Eq.~(\ref{eq:LNP}). 
Here $\alpha$ and $k$ 
denote, respectively, the lepton flavor and NP fermion index. 
}
\label{tab:particles}
\begin{center}
\begin{tabular}{|c||c|c|c|c|c|}
\hline 
&  $SU(2)_{L}$ & $  U(1)_{Y}$   & \ZZ & $S$
\\ \hline \hline 
 $\Phi_{\rm SM}$  & $2$&$ 1$ & $+$ & $0$ \\
\hline
 $\Phi\phantom{_{\rm SM}}$ & $2$&$ 1$  & $-$ & $0$ \\
\hline
 ${L_\alpha}$  & $2$&$-1\phantom{-} $ & $+$ & $1/2$ \\
\hline 
 ${\Sigma}_{k}$ & $3$&$ 0$  & $-$ & $1/2$ \\ 
\hline
\end{tabular} 
\end{center}
\end{table}

The mass ordered  \ZZ-odd fermion fields, triplets of $SU(2)_L$, 
can be written as~\cite{Chao:2012sz}
\begin{eqnarray}\label{eq:Sigma}
{\Sigma_{k}} = \begin{pmatrix}
{\Sigma^{0}_{k} }/{\sqrt{2}} 
& \Sigma^{+}_{k} 
\\[0.2em]
\Sigma^{-}_{k}  & 
- 
{\Sigma^{0}_{k} }/{\sqrt{2}} 
\end{pmatrix}.
\end{eqnarray}
At tree level
the masses for the neutral and charged \ZZ-odd fermion triplets $\Sigma_{k}$ are  
degenerated within each generation.
At one loop the mass splitting 
between the charged and neutral components of $\Sigma_{k}$ 
can be
computed with the general formulae given in Ref.~{\cite{Cirelli:2005uq}, resulting in a mass splitting of between
$\Delta m_{\rm{loop}} \approx 152\mev$
for small  $\msig{k}{0}$,
and 
$
\Delta m_{\rm{loop}}^{\rm{max}} =
\alpha_{2} M_{W} \sin ^{2}  ({\theta_{W}}/{2} ) =
166\pm 1\mev$,
its asymptotic value for large $\msig{k}{0}$.
This mass difference is small enough to neglect decays of the charged fermion to the neutral one and a virtual $W$ boson.

Since our analysis is not sensitive to the $CP$ properties of the model we assume, 
without loss of generality,
that the $CP$-even scalar ${H^{0}}$ is lighter than the $CP$-odd ${A^{0}}$. 
Therefore ${H^{0}}$ is stable and the natural DM candidate.
A convenient set of parameters to describe the full model are the masses of
the unknown scalar spectrum $\lbrace \mdm, \mha, \mhp\rbrace$, the self-couplings 
$\lambda_{2}$, $\lambda_{L} \equiv \lambda_{H^{0}}  =({\lambda_{3} + \lambda_{4} + \lambda_{5}})/2$,
$n_\Sigma\times n_\Sigma$ complex Yukawa couplings $Y_{k\alpha}$, and the $n_\Sigma$ masses for the neutral components
of the fermion triplet $\msig{k}{0}$. 

\subsection{Neutrino Mass Generation}
\label{subsec:neu.mass.gen} 

In this model the neutrino masses arise at one-loop 
via their interaction with the \ZZ-odd fermions and scalars~\cite{Ma:2008cu}. 
The corresponding Feynman diagram is displayed
 in Fig.~\ref{fig:numass}.
The neutrino mass matrix reads 
\begin{align}
\label{eq:Mnuint} 
\nonumber
({\mathcal{M}}_{\nu})_{\alpha\beta} 
&= \sum_{k=1}^{n_\Sigma} Y_{k\alpha} Y_{k\beta}\Lambda_k=\sum_{k=1}^{n_\Sigma} \left[Y^T \Lambda Y\right]_{\alpha\beta}~, \qquad  \alpha,\beta=1,2,3~,
\\
 \Lambda_{k}          
&= \dfrac{\msig{k}{0}}
          {32\pi^{2}}
           \Bigg[ \dfrac{\mdm^{2}}{\mdm^{2}-\msig{k}{0}^{2}}
                  \ln\bigg(\dfrac{\mdm^{2}}{\msig{k}{0}^{2}}\bigg) 
                 - \dfrac{\mha^{2}}{\mha^{2}-\msig{k}{0}^{2}}
                  \ln\bigg(\dfrac{\mha^{2}}{\msig{k}{0}^{2}}\bigg) \Bigg]
~,
\end{align}
where $\Lambda_{k}$ are the entries of the diagonal matrix $\Lambda$. 
The special case $n_\Sigma = 2$ leads to a singular neutrino mass matrix with one vanishing eigenvalue.
The physical neutrino masses are obtained diagonalising Eq.(\ref{eq:Mnuint})
with the Pontecorvo-Maki-Nakagawa-Sakata neutrino mixing matrix $ U_{\rm PMNS} $~\cite{Maki:1962mu}
(see Ref.~\cite{Beringer:1900zz} for its standard parametrization):
\begin{eqnarray}\label{eq:Mnudiag} 
  U_{\rm PMNS}^{T} {\mathcal{M}}_{{\nu}} U_{\rm PMNS} = \text{diag}(m_{\nu_e},m_{\nu_{\mu}},m_{\nu_{\tau}})    
  \equiv M_{\nu}^{\rm diag}  ~.
\end{eqnarray} 
Using the Casas-Ibarra parametrization procedure~\cite{Casas:2001sr}
we express the Yukawa coupling matrix  in terms of the new physics mass parameters
 included in  $\Lambda_{k}$~(\ref{eq:Mnuint}), 
and the experimental neutrino data:
\begin{eqnarray}\label{eq:CasasIbarra}
Y = \sqrt{\Lambda}^{-1} R \sqrt{ M_{\nu}^{\prime \rm diag}}U_{\rm PMNS}^{\dagger}
~,
\end{eqnarray}
\begin{figure}[t]
\begin{centering}
\includegraphics[scale=0.8]{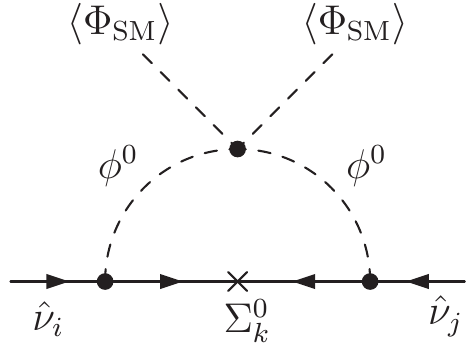}
\caption{\label{fig:numass} 
One-loop neutrino mass generation in the RSIII via the exchange of a \ZZ-odd neutral scalar
$\phi^{0}=H^{0},A^{0}$ and a \ZZ-odd fermion $\Sigma_{k}^{0}$.  
$\hat\nu_{\alpha}$, $\hat\nu_{\beta}$ denote neutrino interaction eigenstates.}
\end{centering}
\end{figure}
where $R$ is an arbitrary ${n_{\Sigma}} \times 3$ orthogonal matrix 
connecting \ZZ-odd fermion and lepton flavor space
and 
$M_{\nu}^{\prime \rm diag} ={\rm diag}( m_{\nu_{1}}, m_{\nu_{2}}, m_{\nu_{3}} )$.
If $n_{\Sigma}>2$ and the lightest neutrino is allowed to vary in its full experimentally allowed range
both hierarchies cover almost the whole range of normalized Yukawa couplings,
as can be observed in Fig.~\ref{fig:NHIH}, where solutions 
of Eq.~(\ref{eq:CasasIbarra}) with real $R$
are shown in flavor space for the normal (NH) and inverse (IH) hierarchies.
Here $ \hat Y_{\alpha }  \equiv\hat Y_{1\alpha} = {Y_{1\alpha}}/{ \sqrt{\sum_{\alpha=e,\mu,\tau} |Y_{1\alpha}|^2 }} $
denote the normalized Yukawa couplings
and the color shows the logarithmically averaged mass of lightest neutrino mass in each hierarchy. 
These solutions have been obtained for ${\Sigma_k^\pm}$ masses of $500, 1500$, and $2500\gev$.
However, qualitatively similar solutions are obtained for different fermion masses.
For our numerical analysis we will assume $n_{\Sigma}=3$ and a normal hierarchy for the neutrino masses. 
\medskip

\begin{figure}[h]
\begin{centering}
\hspace*{-.9cm}
\includegraphics[scale=0.37]{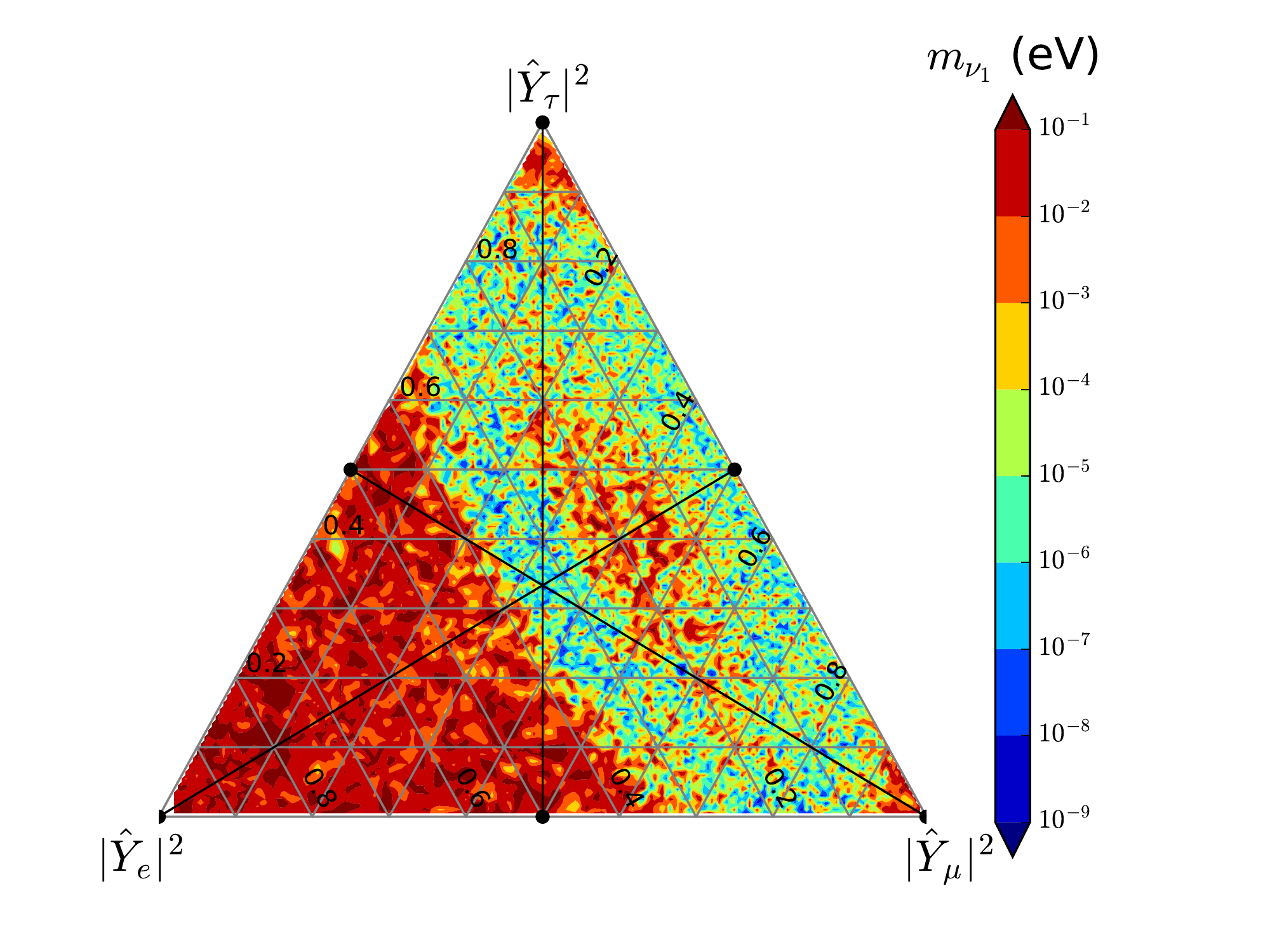}
\includegraphics[scale=0.37]{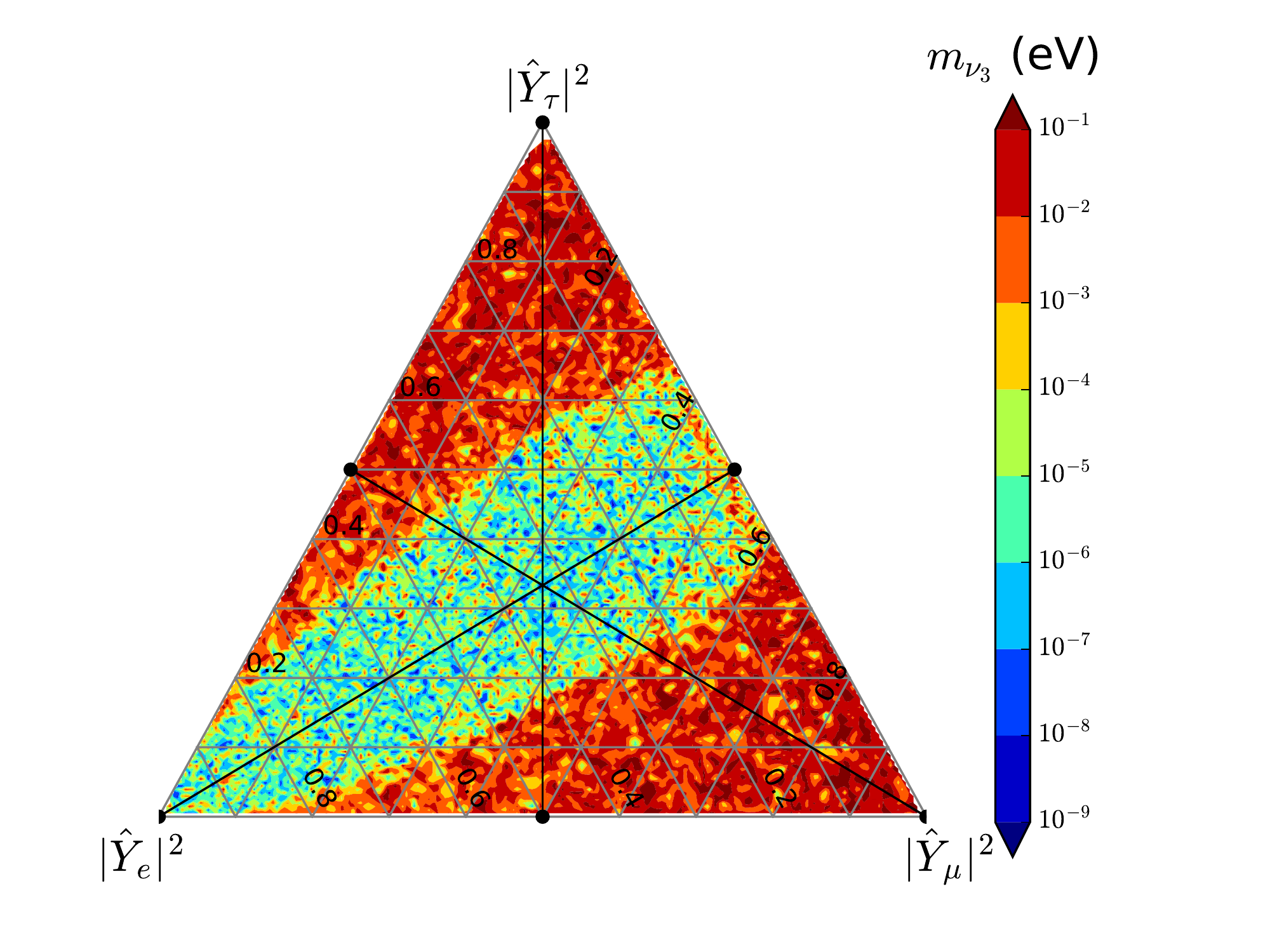}
\put(-405,148){{(NH)}}
\put(-190,148){{(IH)}}
\caption{\label{fig:NHIH}
Normal (NH) and inverse (IH) hierarchy solutions in flavor space (visualized as described in~\cite{Anandakrishnan:2014pva}).
For every set of normalized Yukawa couplings squared $|\hat Y_{\alpha}|^{2}, \alpha=e,\mu,\tau$,
the lightest neutrino mass 
 $m_{\nu_{k}}$ of the obtained solutions is averaged logarithmically. 
}
\end{centering}
\end{figure}

\subsection{Lepton Flavor Violation (LFV)}
\label{subsec:mueg} 

The LFV processes such as $\mu^- \to e^- \gamma $
vanish in the SM but arise in the RSIII at the one-loop 
via the LFV Yukawa interactions with the \ZZ-odd scalars~(\ref{eq:LNP}) shown in Fig.~\ref{fig:mueg}. 
The analytic expression for $Br (\mu^- \to e^- \gamma)$ is given by 
\begin{align}
\label{eq:mueg}
Br (\mu^- \to e^- \gamma)
&=\frac{3 \alpha_{\rm em}Br (\mu^- \to e^- \nu_\mu\overline{\nu}_e)}{256 \pi^2 G_F^2} 
\nonumber\\[.3em]
& \times
\left|
  \sum_{k=1}^{n_\Sigma} Y_{k\mu}^* Y_{k e}
  \left\{
      \frac{1}{\mhp^2} F_{2}\left(\frac{\msig{k}{0}^2}{\mhp^2}  \right)
     -\frac{1}{\msig{k}{\pm}^2} 
     \left[
       F_{2}\left(\frac{\mdm^2}{\msig{k}{\pm}^2}\right)
      +F_{2}\left(\frac{\mha^2}{\msig{k}{\pm}^2}\right)
     \right]
  \right\}
\right|^2
~,
\end{align}
with $G_F$ the Fermi constant and
\begin{align}
F_{2} (x)
&=\frac{1-6x + 3x^2 + 2x^3 -6x^2\log{x} }{6 (x-1)^4}
~.
\end{align}
This expression can be trivially generalized to $\tau^- \to \mu^- \gamma$ and  $\tau^- \to e^- \gamma$.

\begin{figure}[ht]
\begin{centering}
\vspace*{.5cm}
\includegraphics[scale=0.8]{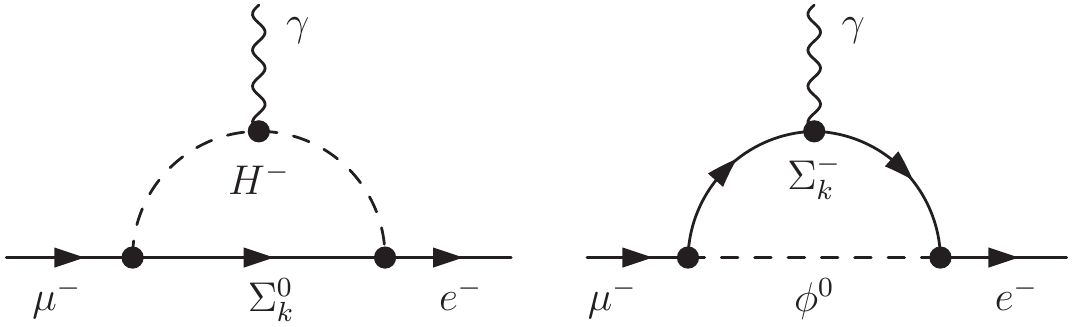}
\caption{\label{fig:mueg} 
Feynman diagrams contributing to $\mu^-\to e^- \gamma$.
Here $\phi^{0}=H^{0},A^{0}$. 
Not shown are the self-energy corrections leading to electron-muon mixing.}
\end{centering}
\end{figure}

\subsection{Dark matter}
The case of fermionic DM  has been studied in~\cite{Ma:2008cu,Chao:2012sz}.
The DM candidate is the neutral component of the lightest NP fermion triplet.
Since its electroweak couplings to gauge bosons are unsuppressed
the \ZZ-odd fermions need to be heavier than around $2.6\tev$~\cite{Chao:2012sz}
in order to suppress the DM annihilation cross-section before freeze-out and thus allow for the correct relic density.
Therefore one does not expect significant phenomenological signatures at the LHC.

The scalar sector, on the other hand, 
 allows for  lighter DM.
Since it has the same field content and couplings to the SM 
 as the Inert Higgs Doublet Model~\cite{Deshpande:1977rw},
its phenomenology is also very similar. 
Scalar DM is viable both at around the electroweak scale, the ``low mass region'', 
as well as above $500\gev$~\cite{Arhrib:2013ela}.
We focus our analysis in the phenomenologically more interesting low mass region for DM. 
In this region
the dominant annihilation channels are $b\bar b$, 
mainly via the exchange of a Higgs boson in s-channel, 
and annihilation to gauge boson pairs for very small $\lambda_L$, or above the $W^+W^-$ threshold.

\section{Model Constraints}
\label{sec:Model_Constraints}

The RSIII model is constrained by direct and indirect searches for DM, colliders and electroweak precision observables.
In this section we review the implication of these constraints on the parameter space of the model.

\subsection{Theory constraints}
\label{sec:th_constraints}
The following conditions are obtained requiring that the scalar potential is bounded from below~\cite{Branco:2011iw}: 
$\lambda_{1,2} > 0$, $\lambda_{3} + \lambda_{4} -|\lambda_{5}| + 2\sqrt{\lambda_{1}\lambda_{2}} > 0$ and $\lambda_{3} + 2\sqrt{\lambda_{1}\lambda_{2}} > 0$. 
Requiring perturbativity 
sets bounds on the scalar couplings, $|\lambda_i| < 8\pi$, for $ i=1,\dots,5$.
However, tree-level unitarity constraints~\cite{Kanemura:1993hm,Arhrib:2012ia}  
set stronger bounds on these couplings (see Ref.~\cite{Arhrib:2012ia}).

\subsection{\label{subsec:EWPO}Electroweak precision observables}
The contribution to the oblique parameters $S$, $T$, $U$ from the \ZZ-odd 
scalar sector 
have been computed for the IDM in Refs.~\cite{Barbieri:2006dq,Baak:2011ze,Okada:2014qsa}.
The contribution to $S$, $T$ and $U$ from the \ZZ-odd fermions, a triplet of $SU(2)_L$, vanish. 
As in the case of pure gauginos in the MSSM, they cannot contribute to 
operators with $SU(2)_L$-breaking quantum numbers, 
see e.g.,~\cite{Barbieri:2004qk,Cynolter:2008ea,Marandella:2005wc}.
The SM best fit obtained in~\cite{Baak:2014ora} 
with a reference SM  defined fixing $m_{t,\rm ref}=173\gev$ and $M_{H,\rm ref}=125\gev$
is
\begin{align}
&
\bar S=0.05\pm 0.11,\quad 
\bar T=0.09\pm 0.13, \quad
\bar U=0.01\pm 0.11,
\nonumber\\
&
\rho_{ST}=+0.90,\quad
\rho_{SU}=-0.59,\quad
\rho_{TU}=-0.83~,
\label{eq:ST_SM}
\end{align}
from which the correlation matrix is computed.

\subsection{Collider constraints}

LEP sets limits on the masses of all charged particles which can be directly produced,
as well as on particles produced as their decay products.
These limits can be easily reinterpreted for the new scalars and fermions of the RSIII. 
 The decays of gauge bosons into \ZZ-odd pairs  
are excluded by their invisible width measurements~\cite{Beringer:1900zz}, 
leading to the constraints $m_{H^{0},A^0}+\mhp> M_{W}$, $\mha+\mdm > M_{Z}$, $2\mhp > M_{Z}$, $\msig{1}{0} + \msig{1}{\pm} > M_{W}$ and $2 \msig{1}{\pm} > M_{Z}$.
Since the \ZZ-odd fermions couple to gauge bosons with the same couplings as the gauginos
we can apply the bounds on direct chargino searches at LEP II $ m_{\Sigma^{\pm}_{1}} > 103.5\gev~$%
~\cite{LEPSUSYWG,Heister:2003zk,Abdallah:2003xe,Acciarri:1999km,Abbiendi:2003sc}.
Direct chargino searches at LEP II
can also be reinterpreted for the search of charged scalars~\cite{Pierce:2007ut}, leading to $ \mhp > 70\gev$. 
The direct LEP search limits for associated scalar and gauge boson  
do not apply here due to the existence of the \ZZ\ symmetry.
We use the bounds obtained in~\cite{Lundstrom:2008ai}
\begin{eqnarray}\label{eq:AHtoRS3.M} 
\label{eq:LEPmA}
{\rm{max}}(\mha,\mdm) \gsim 100{\gev}
\qquad {{\rm or}}
\qquad |\mha - \mdm|< 8\gev
~.
\end{eqnarray} 
Since the bound on the heavier neutral scalar varies between $100\gev$ and $110\gev$ 
as a function of the lightest scalar mass (see Fig.~7 of Ref.~\cite{Lundstrom:2008ai}). 
We require ${\rm{max}}(\mha,\mdm) > 110{\gev}$.
The small allowed region for ${\rm{min}}(\mha,\mdm) \ge m_W$ and $|\mha - \mdm|> 8\gev$ which we exclude 
does not significantly affect our analysis.

The LHC sets bounds on the invisible and diphoton Higgs decays. 
If any of the channels 
 $h \to H^{0}H^{0}, A^{0}A^{0}$, 
are open, they should satisfy the constraint on the upper limit for the invisible decay of the Higgs boson 
~\cite{ATLAS-CONF-2015-044}
 \begin{align}
 \label{eq:BRhinv}
 & 
 \sum_{\Phi^{0} = H^{0}, A^{0}} {\rm Br}(h \to \Phi^{0} \Phi^{0}) < {\rm{Br}}^{\rm max}(h \to \rm {inv.}) 
 = 0.13~. 
 \end{align}
This upper limit is expected to be reduced by half 
at the future Run-II of the LHC~\cite{Abe:2014gua}.  
For the diphoton channel, the signal strength $R_{\gamma \gamma}$ measures the 
ratio of the observed diphoton production cross section relative to the SM expectation~\cite{Posch:2010hx}: 
\begin{align}
 \label{eq.hgaga}
R_{\gamma \gamma} &= \dfrac{\sigma(pp \to h \to \gamma \gamma)^{\rm{RSIII}}}{\sigma(pp \to h \to \gamma \gamma)^{\rm{SM}}} 
            = \dfrac{\sigma(pp \to h \to \gamma \gamma)^{\rm{IDM}}}{\sigma(pp \to h \to \gamma \gamma)^{\rm{SM}}} 
                          \approx \dfrac{[{\rm Br}( h  \to \gamma \gamma) \big ] ^{\rm{IDM}}}{[{\rm Br}( h \to \gamma \gamma)\big ]^{\rm{SM}}}~.
\end{align}
This relation holds since the \ZZ-odd fermions do not interact with the SM Higgs boson.
The signal strength relative to the Standard Model expectation is
measured by ATLAS~\cite{Aad:2014eha} and CMS~\cite{Khachatryan:2014ira},
\begin{align}
\label{eq:rgagaLHC}
R_{\gamma \gamma}^{\rm ATLAS}  = 1.15\pm ^{+0.27}_{-0.25}~,\qquad
 & R_{\gamma \gamma}^{\rm CMS}  = 1.12\pm ^{+0.25}_{-0.23}~.
\end{align}

\subsection{Flavor  constraints}
\label{sec:neu.osc}

An analysis of LFV in the RSIII has been  carried out in 
Ref.~\cite{Chao:2012sz} for the case of fermionic DM,
where bounds on the Yukawa couplings have been derived.
The results from a recent analysis of LFV processes in the minimal scotogenic model for fermion DM masses of up to 3~TeV~\cite{Vicente:2014wga}
can be extended to the RSIII.
These bounds, however, do not directly apply for our case, with significantly lighter NP fermions.
In our model the  Yukawa couplings, 
which are obtained from the neutrino masses,
 turn out to be at most of order $\mathcal{O}(10^{-4})$ 
if we choose the orthogonal matrix $R$ of Eq.~(\ref{eq:CasasIbarra}) real. 
In this case the LFV bounds do not further constrain the available parameter space. 
On the other hand, if $R$ is allowed to be complex, 
much larger values of the Yukawa couplings can be obtained and the $\mu^+\to e^+\gamma$ and $\tau^{} \to \mu^{} \gamma$
bounds~\cite{Adam:2013mnn,Abe:2003sx} 
restrict their largest values,
of approximately $1$ ($0.5$)
for the electron Yukawa in the normal (inverted) hierarchy,
and of order of a few for the muon and tau Yukawa couplings.

\subsection{Dark Matter constraints}
\label{sec:DMconstr}
The DM relic density 
measured by Planck~\cite{Ade:2015xua} \footnote{We 
have used the result for Planck TT+lowP of Ref.~\cite{Ade:2015xua}.
A tighter bound is given for Planck TT,TE,EE+lowP, which does not significantly alter our analysis.}
in units of the critical density and the normalized Hubble constant $h$
is 
$\Omega_{\rm DM}^{\rm exp.} h^{2} = 0.1197 \pm 0.0022$ at $68\%$~confidence level (CL).
Allowing for other unknown sources for DM this measurement only imposes a upper bound on the NP contribution to $\Omega_{\rm DM}h^2$.
In the numerical analysis 
we require that the relic density lies within a $2\sigma$ uncertainty of the measured central value,
$\Omega_{\rm DM} h^{2} = 0.1197\pm 0.0044 $. 
Whenever we relax this constraint to allow for additional DM sources we only require that $\Omega_{\rm DM} h^{2} < 0.1241 $.

With respect to direct DM searches, 
we use the $90 \% $~CL upper bound of the spin-independent DM-nucleon cross section
$\sigma_{\rm SI}^{\rm max}$
 given by LUX~\cite{Akerib:2015rjg}.    
Allowing for an underabundance of DM this bound is rescaled as
\begin{align}
\label{eq:sigmasi} &
\sigma_{\rm SI} < 
\xi^{-1}_{\rm DM} \sigma_{\rm SI}^{\rm max} ~,
\end{align}
with
 $\xi_{\rm DM}=\Omega_{\rm DM}/\Omega_{\rm DM}^{\rm exp.}<1$
the ratio of the DM relic density of  our model
and the experimental central value obtained by Planck~\cite{Ade:2015xua}.
The lower DM density leads to a smaller sensitivity for direct detection 
and consequently to a larger upper limit on the spin-independent DM-nucleon cross section.
Here one assumes that all remaining unknown sources of DM 
do not contribute to the direct detection signal.

For indirect DM searches, 
we use the $95\%$~CL upper bound of the thermally averaged cross-section 
 obtained by Fermi-LAT~\cite{Ackermann:2015zua}
for dwarf spheroidal galaxies with the 6-year Pass-8 Limit.
In order to account for the different annihilation channels of our DM candidate
we normalize the corresponding bounds for 
$\langle \sigma v  \rangle_{X}$, with $X= b\bar b, WW,ZZ,hh$, 
and select the strongest one. 
Allowing for an underabundance of DM this bound is rescaled as
\begin{align}
\label{eq:sigmav} &
\langle \sigma v  \rangle_{X}
  < \xi_{\rm DM}^{-2} 
\langle \sigma v  \rangle ^{\rm max} ~.
\end{align}

\subsection{Scalar sector}
\label{sec:scalar}
As already discussed in the introduction, 
the scalar sectors of the RSIII and the IDM%
~\cite{Deshpande:1977rw,Barbieri:2006dq,LopezHonorez:2006gr,Dolle:2009fn,Honorez:2010re,LopezHonorez:2010tb,Sokolowska:2011sb,Gustafsson:2012aj}
are the same, with the addition of Yukawa couplings to the \ZZ-odd fermions and leptons.
Therefore the RSIII allows for a suitable scalar DM candidate satisfying all model constraints in two regions: 
the low energy region, with a DM mass below the $W$ gauge boson mass,
and the high energy region, with scalar masses above $500\gev$. 
We focus on the first region, where direct production of the \ZZ-odd fermions with large cross-sections is possible. 
We consider DM masses up to $120\gev$ for the low mass region in order to assess the LHC expectations in the region where the DM relic density is less than the one measured by Planck.
It should be noted, however, that in our numerical analysis of Sec.~\ref{sec:results} 
we only consider scenarios where the DM relic density
corresponds to the observed value measured by the Planck collaboration~\cite{Ade:2015xua}.

\begin{figure}[t]
\begin{centering}
 \hspace*{.1cm}
 \includegraphics[scale=0.7]{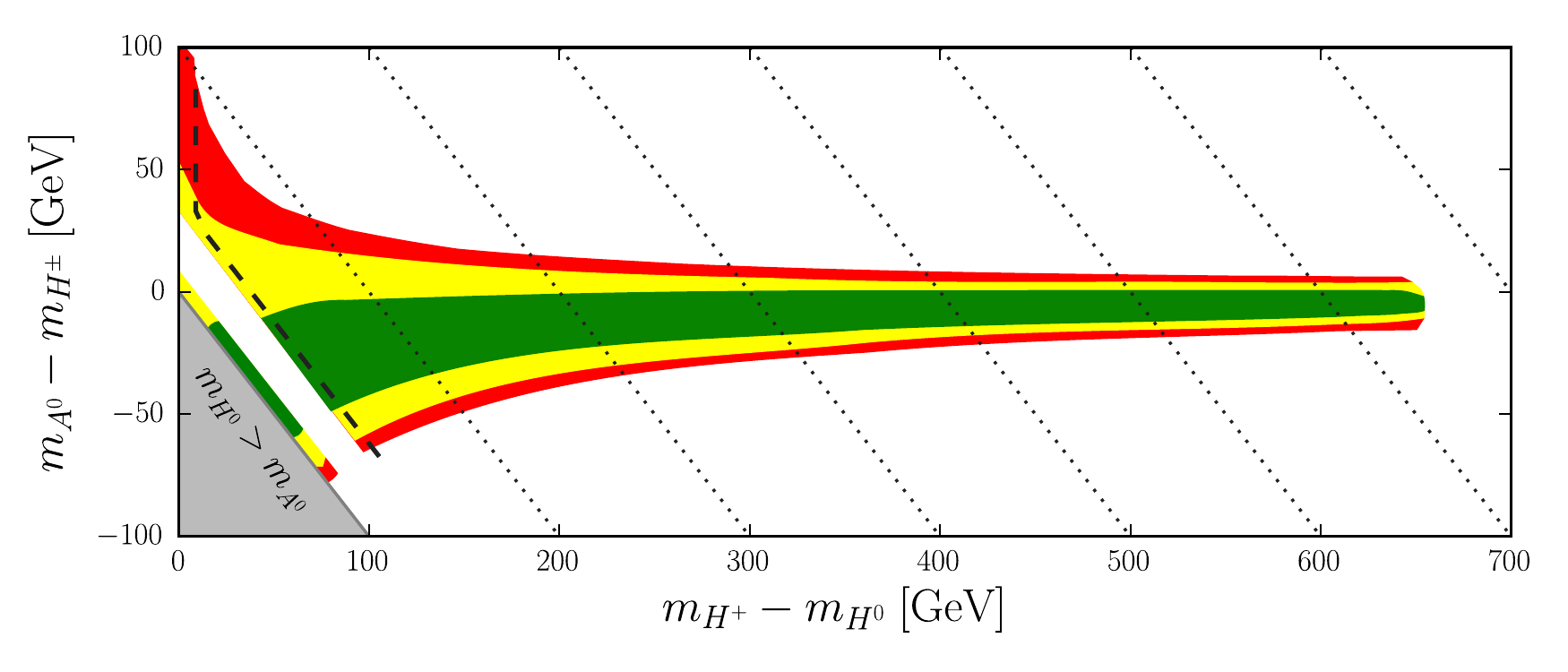}
\caption{\label{fig:MIDM_STU} 
EWPO constraints in the  $(m_{H^{+}} - \mdm )$,$(\mha- \mdm)$ plane.
The regions allowed at $68\%$ (green), $95\%$ (yellow) and $99\%$ (red) CL
have been obtained from the new physics contributions to  the oblique parameters $S,T,U$.
The dashed line delimits from below the region which allows for the correct scalar DM relic density in the RSIII.
All shown points correspond to scenarios which satisfy the constraints of Sec.~\ref{sec:Model_Constraints} for $\mdm<80\gev$.
The gray area corresponds to $\mdm> \mha$, the dotted lines are contours of constant $\mha- \mdm$.
}
\end{centering}
\end{figure}
The constraints from electroweak precision observables (EWPO) strongly restrict the masses of the heavier scalars.
The $\chi$-square for three degrees of freedom, $\chi_{3}^2$, is obtained
from the difference between the 
 oblique parameters $S$, $T$ and $U$, computed following Refs.~\cite{Barbieri:2006dq,Okada:2014qsa,Baak:2011ze,Baak:2014ora},
and their best fit point from EWPO for the SM, Eq.~(\ref{eq:ST_SM}).
In Fig.~\ref{fig:MIDM_STU} we show, for $\mdm$ between $45\gev$ and $80\gev$,
 the  allowed regions at $68\%$ (green), $95\%$ (yellow) and $99\%$ (red)~CL
in the $(\mha-\mhp)$, $(\mhp-\mdm)$ plane, 
corresponding, respectively, to  $\chi^2_{3} \le {3.506}$, $\chi^2_{3} \le 7.815$, and $\chi^2_{3} \le {11.345}$.
The two remaining free parameters of the scalar sector, $\lambda_{2}$ and $\lambda_L$, have no effect on the oblique parameters.
Contours of constant $\mha-\mdm$ are shown
as dotted lines.
The gray area in Fig.~\ref{fig:MIDM_STU} corresponds to $\mdm>\mha$, 
for which $H^{0}$  is not the DM candidate.
The stronger constraints come from $T$, 
which depends on the differences of masses between charged and neutral scalars, 
and $S$, which is sensitive to the difference of the neutral scalar masses.
The dependence on $\mdm$ is weak but 
can be observed as a small overlap between the different CL regions in the low mass region.
It should be noted that 
 the contribution from the parameter $U$ is often neglected, fixing $U=0$
and evaluating the EWPO constraints with two degrees of freedom.
In our case setting $U=0$  leads to slightly narrower $68\%,\ 95\%,\ 99\%$~CL allowed regions.
The difference of the two choices is due to the fact that,
while the central value of $U$ and the contribution from the IDM to $U$ are small, 
the correlation between the oblique parameters $S$, $T$ and $U$ is large (\ref{eq:ST_SM}). 

In the allowed region where $A^{0}$ and $H^{\pm}$ decouple, with $\mha, \mhp\gg \mdm$, the heavy scalars are nearly degenerate.
The upper bound on  $\mha- \mdm$ and $\mhp-\mdm$, of roughly $650\gev$, 
 follows from the perturbativity constraints given in Sec.~\ref{sec:th_constraints}. 
Also shown is the region for which the correct relic density can be obtained in the low mass DM case analyzed here,
delimited to the left by a dashed line,
excluding small mass splittings between the DM candidate and the heavier scalars
(see also the discussion on Fig.~\ref{fig:omega.vs.mh0}).
Scenarios with $\mha-\mdm$ between roughly $8\gev$ and $30\gev$
are further restricted by the LEP constraints on the second lightest neutral scalar, Eq.~(\ref{eq:LEPmA}),
within the range of DM masses considered here.
In our analysis we have set  conservatively $\mha>110\gev$. 

For DM masses above $80\gev$ the allowed range increases.
For instance, for $\mdm=1\tev$ and $\mhp \approx \mdm$, 
the EWPO constrain $\mha-\mhp \approx \mha-\mdm  \lsim 110\gev$ at $95\%$~CL 
instead of approximately $\lsim 50\gev$ as in the low DM mass case.
Requiring in addition for $\mdm>500\gev$ that these scenarios satisfy the measured relic density 
 leads to $\mha-\mdm \lsim 12\gev$ and $\mhp-\mdm \lsim 8\gev$.

\medskip

The constraints on the scalar sector from 
the thermal relic density measurements, direct and indirect detection, as well as the LHC,
are analyzed performing a scan of the following parameters in the range
\begin{align}
\label{eq:param.scan} 
\nonumber
  45{\gev} <&\ \mdm < 120{\gev}~,
\\ \nonumber
110{\gev} <&\ \mha < 700{\gev} , 
\quad  {\rm or} \quad 
0< \mha-\mdm < 8{\gev} ~,
\\ \nonumber   
  70{\gev}   <&\ \mhp <  700{\gev} ,
\\ 
 10^{-5} <&\ |\lambda_{L}| < |\lambda_{L}|^{\rm max} ~,
\end{align}
and fixed $\lambda_2=0.1$. 
The value of $\lambda_2$ is irrelevant for our study, as long as it fulfills the theory constraints. 
We have computed the spin-independent DM-nucleon cross section $\sigma_{\rm SI}$, 
the thermal averaged annihilation cross-section $\langle \sigma v  \rangle $~(\ref{eq:sigmav}),
and $R_{\gamma \gamma}$~(\ref{eq.hgaga}) with the IDM model of {\tt micrOMEGAs (v4.1.8)}~\cite{Belanger:2013oya}.
We have confirmed these results comparing  $\sigma_{\rm SI}$ and $\langle \sigma v  \rangle $ with Ref.~\cite{Arhrib:2013ela},
and  $R_{\gamma \gamma}$ following the treatment carried out in Ref.~\cite{Swiezewska:2012eh}.
We also impose the EWPO, perturbativity 
of Sec.~\ref{sec:Model_Constraints}.
The choice of parameters also satisfies the LEP collider constraints~\cite{Lundstrom:2008ai}.
The value of
 $ |\lambda_L|^{\rm max} $ depends on the specific parameter point 
and is obtained from the perturbative unitarity constraint. 
All values of $\mha $ and $ \mhp $ are below their perturbativity limit.
The fermion sector has no effect on the DM observables due to the smallness of the Yukawa couplings%
\footnote{\label{comment.complexR}
We have restricted our analysis to the case of real orthogonal matrix $R$ (\ref{eq:CasasIbarra}).
The solutions with large Yukawa couplings obtained allowing $R$ to be complex are highly fine-tuned~\cite{Ibarra:2010xw}.
The compatibility with the neutrino oscillation data~\cite{Forero:2014bxa}, achieved through Eq.~(\ref{eq:CasasIbarra}),
receives large higher order radiative corrections~\cite{AristizabalSierra:2011mn}
which spoil the fine-tuning obtained at leading order. 
It is worth noticing that, while new DM annihilation channels may become significant, 
allowing for new lighter scalar DM solutions, 
 the experimental signatures from heavier \ZZ-odd fermion decays should not significantly modify our phenomenological analyisis,
as can be inferred from SUSY searches.
}
}, 
not larger than $\mathcal{O}(10^{-4})$. 

\begin{figure}[t]
\begin{centering}
  \hspace*{-0.75cm}
\includegraphics[scale=0.66]{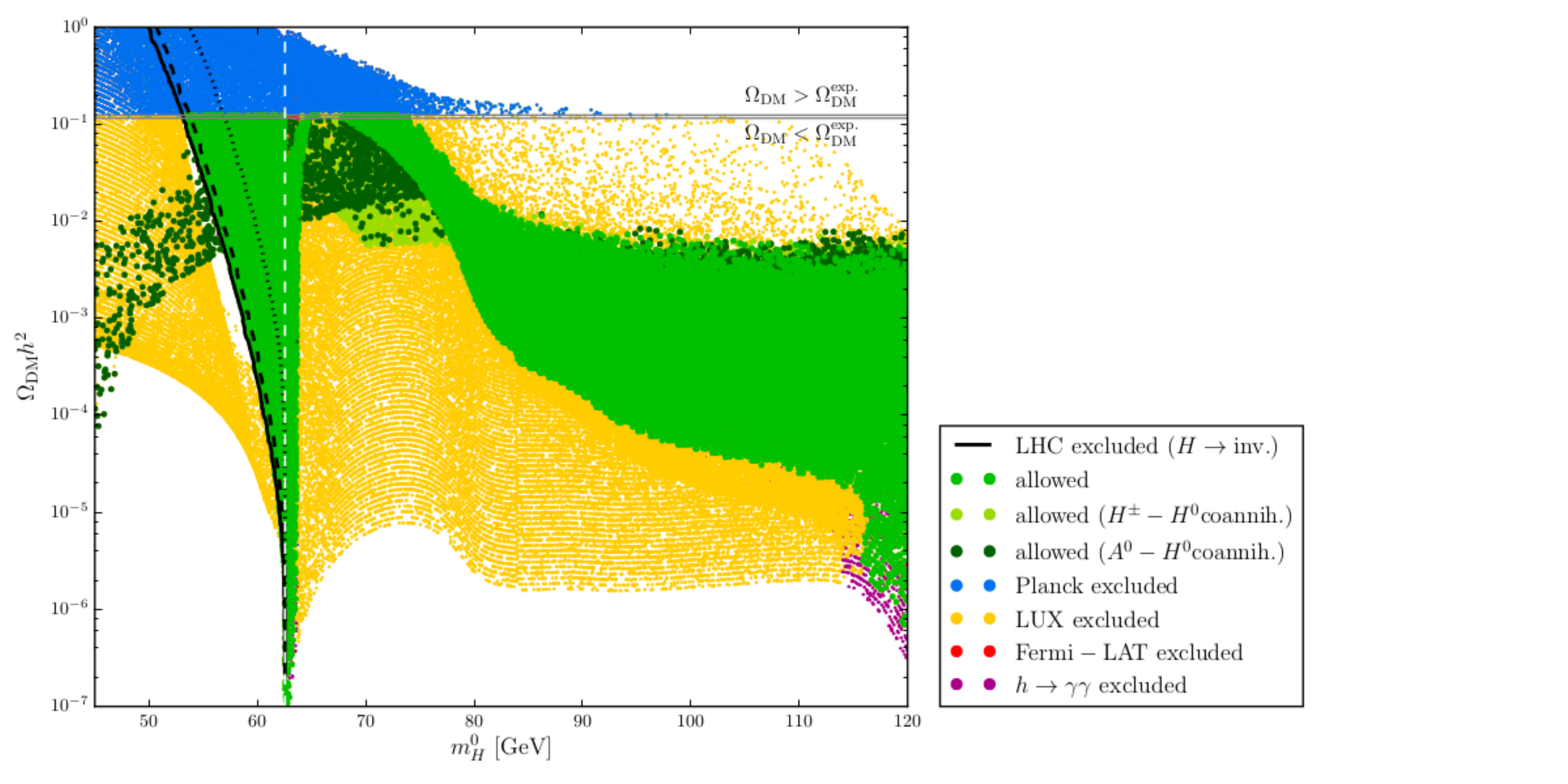}  
\caption{
\label{fig:omega.vs.mh0} 
Constraints of the scalar sector of the RSIII  
in the $\Omega_{\rm DM}h^2,\mdm$ plane.
Green points satisfy all constraints of Sec.~\ref{sec:Model_Constraints}. 
Dark green (light green) points represent  
scenarios with $\mha-\mdm<8\gev$ ($\mhp-\mdm<12\gev$),  
in which $A^{0}$--$H^{0}$ ($H^{\pm}$--$H^{0}$) 
co-annihilation is the dominant annihilation channel before freeze-out. 
The upper bound on the invisible Higgs decay width from the LHC (black curve)
gives a lower bound on $\mdm$ except for the {dark green points ($H^{0}$--$A^{0}$ co-annihilation scenarios)}. 
The remaining scenarios are excluded by Planck relic density measurement (light blue),
 LUX direct detection searches (yellow),
Fermi-LAT indirect detection searches (red),
LHC Higgs decay to photons (purple).
The bound for the invisible Higgs decay
for a naive projection at the LHC Run-II (ILC expected sensitivity)
is shown as a dashed (dotted) black curve.
The horizontal lines represent the $2\sigma$ band on the measured relic density.
A vertical dashed  gray line shows the threshold of Higgs decay to DM pairs. 
}
\end{centering}
\end{figure}
The result of the scan of parameters is shown in Fig.~\ref{fig:omega.vs.mh0}
in the relic density versus DM mass plane.
Scenarios which fulfill all constraints are shown as green, dark green and light green dots. 
Dark green (light green) dots represent scenarios in which the main annihilation channel before freeze-out
is the co-annihilation between $A^{0}$ and $H^{0}$ ($H^{\pm}$ and $H^{0}$),
defined here by $\mha-\mdm < 8\gev$ ($\mhp-\mdm < 12\gev$).
The mass difference between the coannihilating scalars is small enough to avoid the Boltzmann suppression before freeze-out. 
As this mass splitting increases, the annihilation cross-section decreases, 
leading to a larger relic density.
For instance, at low $\mdm$ the $A^{0}$--$H^{0}$ co-annihilation scenarios have a lower limit in $\Omega_{\rm DM} h^2$ 
when the splitting vanishes, 
and an upper limit when it reaches its maximum value of $8\gev$,
 implying that for low DM masses the coannihilation mechanism 
is too efficient to allow for the observed relic density. 
The dark green dots in the light green region for $\mdm \approx 70\gev$ 
correspond to scenarios where both heavier scalars coannihilate with $H^{0}$.
For larger values of  $\Omega_{\rm DM} h^2$ both co-annihilation regions overlap 
but the dark dots cover the light ones. 
Similarly, the green dots cover the light and dark ones where those regions overlap.

Scenarios excluded by the upper bound on the relic density measurement by Planck~\cite{Ade:2015xua}, Sec.~\ref{sec:DMconstr}, are shown in light blue.
Scenarios with a smaller value of $\Omega_{\rm DM}h^2$ are not excluded but lead to an underabundance of DM 
which cannot fully account for the DM content of the Universe.
In that case the  direct detection upper bound on the spin independent cross-section $\sigma_{\rm SI}^{\rm max} $ 
is rescaled with $\xi^{-1}_{\rm DM}$ as in Eq.~(\ref{eq:sigmasi})
to take into account the smaller DM flux on the detector.
Analogously, the indirect detection upper bound on the thermally averaged cross-section 
is rescaled with  $\xi^{-2}_{\rm DM}$ as in Eq.~(\ref{eq:sigmav}).
The upper bound on the relic density excludes scenarios without an efficient mechanism of annihilation before freeze-out.
These scenarios are characterized by a large splitting between $H^{0}$ and the heavier scalars, 
suppressing the co-annihilation channels,
and, for $\mdm<M_W$  a small DM--Higgs coupling $\lambda_L$, suppressing the Higgs exchange channel, 
while for $\mdm \approx M_W$, by $\lambda_L \sim {\mathcal{O}}(-0.1)$, 
leading to a destructive interference between different annihilation channels to gauge bosons.
Also shown are the maximum and minimum allowed values for the relic density 
as measured by Planck at $95\%$~CL level if one requires that the model fully explains the DM content of the Universe.

The strongest constraint from the LHC comes from 
the present bound on the invisible branching ratio of the Higgs boson, 
shown as a black solid line,
which sets a lower mass limit for $H^0$ 
whenever the Higgs-portal is the main DM annihilation channel.
For $\xi_{\rm DM}=1 $ this bound excludes $\mdm<53\gev$. 
For $\xi_{\rm DM}<1 $ it excludes scenarios with masses 
of up to $m_h/2$,
unless the $H^{0}$--$A^{0}$ co-annihilation channel contributes significantly to the total annihilation before freeze-out.
In the later case, corresponding to the band of dark green points in the light DM mass region,
the DM-Higgs boson coupling $\lambda_L$ is small enough to restrict the invisible Higgs decay,
while the co-annihilation channel ensures that the Planck upper limit on the relic density is fulfilled.
Also shown as a black dashed line
is the future projection of the upper limit on the invisible decay of the Higgs boson at Run-II of the LHC 
assuming a future limit for the invisible Higgs decays 
${\rm Br}^{\rm LHC13}(h\to {\rm inv.})<0.065$~\cite{Abe:2014gua}, 
and  as a black dot-dashed line the corresponding prospect for the ILC
with $\sqrt{s}=1\tev$ and $1~{\rm ab}^{-1}$~\cite{Baer:2013cma}, ${\rm Br}^{\rm ILC}(h\to {\rm inv.})<0.0026$.

Scenarios allowed by Planck upper limit but excluded by the direct detection constraints from LUX~\cite{Akerib:2015rjg}  
are shown in yellow.
The lower sensitivity to the spin independent cross-section resulting  
when the relic density is smaller than the experimental measured value, 
obtained rescaling the upper limit with the factor the $\xi_{\rm DM}^{-1}$,
reduces the excluded region significantly.
The direct detection limit also depends on variations on the local DM density, 
which would have to be included in the factor  $\xi_{\rm DM}$. 
It is interesting that, 
for $\Omega_{\rm DM} = \Omega_{\rm DM}^{\rm exp.} $, 
the lower bound on $\mdm$ from LUX is only slightly stronger than that from the invisible Higgs decay. 
For $\Omega_{\rm DM} = \Omega_{\rm DM}^{\rm exp.} $
LUX also sets the upper limit $\mdm<74\gev$,
corresponding to scenarios with $\lambda_L \approx {-0.012}$.
Larger values of $\mdm$ require larger values of $|\lambda_L|$ in order to obtain the correct relic density, 
increasing the spin independent cross-section above the LUX bound.
Allowing for DM underabundance LUX constrains regions of parameter space up to $\mdm=120\gev$. 
For $\mdm>110\gev$ and  $\lambda_L\ne 0$ the Higgs pair-production channel becomes a relevant annihilation channel, 
further reducing the relic density and relaxing the constraints due to the rescaling of the bounds.

The indirect detection constraint from Fermi-LAT~\cite{Ackermann:2015zua}, shown in red, 
does not exclude any region of parameter space allowed by the relic density upper limit~\cite{Ade:2015xua}
after we rescale the thermally averaged cross-section by  $\xi_{\rm DM}^{-2}$. 
A small region with $\Omega_{\rm DM} \approx \Omega_{\rm DM}^{\rm exp.} $ and  $\mdm\gsim m_h/2$, in the funnel region, 
is only allowed if the splitting between $A^{0}$ and $H^{0}$ is small and the co-annihilation channel opens up before freeze-out. 

Once all DM constraints are imposed
the LHC measurement of the ratio of the observed diphoton production cross section 
relative to the SM expectation~\cite{Posch:2010hx} constrains a small region of the parameters 
with $\mdm\gsim 114\gev$ and a very small value of relic density. 

For ${120\gev}<m_{H^0}<{500\gev}$, 
where the model leads to an underabundance of DM,
the Higgs diphoton decay restricts a small region in relic density versus DM mass plane with very small relic density, 
corresponding to large $\lambda_L$ and light $H^\pm$.

\section{Phenomenology} 
\label{sec:pheno}
In this section we analyze the phenomenological implications of the constraints on our model given in Sec.~\ref{sec:Model} 
in order to select representative benchmark scenarios for LHC searches.
\medskip

Although the
\ZZ-odd fermion sector of the RSIII 
has the same gauge quantum numbers as
the Type III Seesaw model~\cite{Foot:1988aq}, 
the limits obtained for the latter 
by ATLAS~\cite{Aad:2015cxa} and 
CMS~\cite{CMS-PAS-EXO-14-001,CMS-PAS-EXO-16-002} 
cannot be interpreted as limits in our model 
due to its \ZZ\ symmetry, which forbids the decay of the \ZZ-odd fermions to SM particles. 

\begin{figure}[t]
\begin{centering}
\hspace{-.5cm}
\includegraphics[scale=0.9]{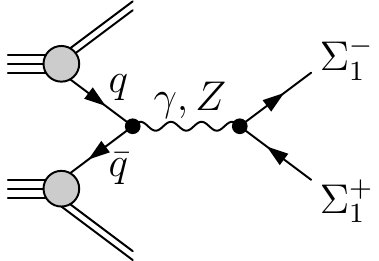}
\hspace*{2cm}
\includegraphics[scale=0.9]{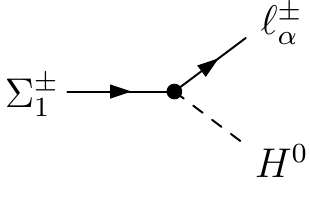}
\put(-190,80){{(a)}}
\put(-40,80){{(b)}}
\caption{\label{fig:FD.prod.dec} 
The left panel~(a) shows the main production channel for pair $\Sigma_1^-\Sigma_1^+$ at the LHC.
The right panel (b) shows the main decay channels of $\Sigma_1^\pm$ to DM.
Here $q$ denotes quarks of the first generation and $\ell_\alpha=e,\mu,\tau$.
}
\end{centering}
\end{figure}
The main production channel of lighter \ZZ-odd fermions at the LHC is shown in  Fig.~\ref{fig:FD.prod.dec}a.
At the LHC gauginos are produced via the s-channel exchange of a gauge boson and via t-channel exchange of a left-handed squark.
Since the gauge structure of the \ZZ-odd fermions and that of charginos and neutralinos in the pure gaugino limit is the same,
their gauge couplings are also equal. 
Therefore,  the production cross-section of \ZZ-odd fermions at the LHC can be obtained 
from that of charginos and neutralinos in the pure gaugino limit with decoupled sfermions,
where the t-channel can be neglected.
For large values of the supersymmetric Higgsino parameter $\mu$
we have checked that the Higgsino component of the chargino is negligible 
and that the results are independent of its value.
We restrict our analysis to the lightest family,  $\Sigma_{1}$, 
for which one obtains the largest production cross-section of \ZZ-odd fermion pairs, 
$  p p \to \Sigma_{i}^{+} \Sigma_{i}^{-} $, $i=1,\ldots,n_\Sigma$.
Our conclusions should be easily extended to the heavier \ZZ-odd fermions.
Notice that 
two-body decays from the heavier \ZZ-odd fermions to the lighter ones are forbidden because the mixing mass matrix $M_\Sigma$ is diagonal.

At tree level the \ZZ-odd fermions decay via Yukawa interactions to a \ZZ-odd scalar and lepton. 
The Yukawa couplings are obtained varying the free neutrino parameters and applying the Casas-Ibarra prescription, 
Eq.~(\ref{eq:CasasIbarra}). 
In the simplest scenario only $H^{0}$ is lighter than the fermion,                              
with the heavier scalars $A^{0}$ and $H^{\pm}$ decoupled and nearly degenerate.
In this case, shown in Fig.~\ref{fig:FD.prod.dec}b,
 both fermions decay exclusively to a lepton and the DM candidate, 
\begin{equation}
\label{eq:sig2l} 
 \Sigma_{1}^{\pm} \to \ell_\alpha^{\pm} H^{0} \quad (\ell_\alpha=e,\mu,\tau)~,
\end{equation} 
resulting in final state dileptons plus MET. 
This channel is expected to be the ``best case scenario'' for \ZZ-odd fermion searches at the LHC.
Neglecting the lepton masses the branching ratios for the decay of the \ZZ-odd lepton 
are proportional to the absolute square of the normalized Yukawa couplings,  
\begin{equation}
\label{eq:Br}
\mathcal{B}_\alpha \equiv {\rm Br} (\Sigma_{1}^{\pm} \to \ell_\alpha H^{0}) = |\hat Y_\alpha|^2~.
\end{equation}
The \ZZ-odd fermion pair-production channel with the  largest production cross-section is $\Sigma_{1}^{0} \Sigma_{1}^{\pm}$.
However,
$\Sigma_{1}^{0}$ decays exclusively to the invisible final state $\nu_\alpha H^{0}$, 
leading to a final state with only one charged lepton and will not be considered here.
Notice that in the Type III Seesaw model, which has the same fermionic content, the decay chains are different 
 due to the absence of a discrete symmetry, leading to different collider signatures~\cite{Aguilar-Saavedra:2013twa}.

If more than one scalar is lighter than the \ZZ-odd fermion,
new decay channels to unstable particles open up, 
\begin{align}
\label{eq:sigmacascades} 
 \Sigma_{1}^{\pm} \to \ell_\alpha^{\pm} A^{0},\quad \Sigma_{1}^{\pm} \to \nu_{\beta} H^{\pm},\quad 
 (\ell_\alpha=e,\mu,\tau;\ \nu_\beta=\nu_e,\nu_\mu,\nu_\tau)~,
\end{align}
followed by the secondary decays
\begin{align}
\label{eq:sigmacascades2} 
& A^{0} \to H^{0} Z,\quad H^{\pm}  \to H^{0} W^{\pm}~,
\end{align}
as well as the subleading decays $A^{0} \to H^{\pm} W^{\mp} $ or $ H^{\pm} \to A^{0} W^{\pm} $.
The gauge boson of the secondary decays may be on-shell or virtual, depending on the mass spectrum.
In addition, the $\Sigma_{1}^{\pm} \Sigma_{1}^{0} $ production channel may lead to final states with at least two leptons,
of either opposite sign or same sign,
\begin{align}
\label{eq:sigmacascades3} 
&  p p \to \Sigma_{1}^{\pm} \Sigma_{1}^{0}, \quad
 \Sigma_{1}^{\pm} \to \ell_\alpha^{\pm} A^{0}/H^{0},\quad  \Sigma_{1}^{0} \to {\ell_\beta^{\mp}} H^{\pm} \quad (\ell_\alpha,\ell_\beta =e,\mu,\tau)
~,
\end{align}
followed by the secondary decays of Eq.~(\ref{eq:sigmacascades2}).
Not shown in (\ref{eq:sigmacascades3}) are the decays to a neutrino and a scalar.
The partial decay width of the decays of Eq.~(\ref{eq:sigmacascades3}) is given by 
\begin{align}
\Gamma (\Sigma_1 \to \ell_\beta \Phi^0) 
&=
\frac{|Y_{1\beta}|^2}{{64 \pi}}\frac{(m_{\Sigma_1}^2 - m_{\Phi^0}^2)^2}{m_{\Sigma_1}^3},\quad \Phi^0=H^0,A^0~,
\\
\Gamma (\Sigma_1^0 \to \ell_\beta^\pm H^\mp) 
&=
\frac{|Y_{1\beta}|^2}{{32 \pi}}\frac{(m_{\Sigma_1}^2 - m_{H^\pm}^2)^2}{m_{\Sigma_1}^3}~.
\end{align}
If all scalars are lighter than $\Sigma_{1}^{\pm}$ and nearly degenerate
the branching ratios
for $\Sigma_{1}^{\pm}$ decaying to $H^{0}$, $A^{0}$ and $H^{\pm}$ 
tend to the asymptotic values $1/4$, $1/4$ and $1/2$, respectively.

\subsection{Collider limits}
\label{sec:pheno_coll} 
Processes with electroweak pair-production and decay of \ZZ-odd particles at colliders, and in particular at the LHC, 
have been extensively studied in the framework of supersymmetry.
Those searches can be interpreted in the framework of the RSIII to constrain this model.
The pair-produced \ZZ-odd particles cascade further to the LOP,
leaving similar collider signatures as those searched for. 
The most convenient way to analyze those results are simplified model spectra analyses,
where  limits on the production cross-sections for NP searches are given as a function of the spectrum.

We focus on a set of benchmark scenarios with well defined decay topologies
and compare these results to LHC searches for supersymmetric processes.
The simplest decay topology is that 
in which both \ZZ-odd fermions decay to the DM candidate, Eq.~(\ref{eq:sig2l}), 
leading to a collider signature of hard opposite sign leptons plus MET. 
Both slepton and chargino pair-production and decay can lead to similar final state topologies.
Pair-production of left-handed sleptons, 
where each slepton decays further to the lightest neutralino and a lepton of the first two families,
$ p p \to \tilde{\ell}_{L}\tilde{\ell}_{L} \to \ell^{\pm} \ell^{\mp} \tilde{\chi}_{1}^{0} \tilde{\chi}_{1}^{0} $, with $\ell=e,\mu$
and $\widetilde{\chi}_{1}^{0}$ the lightest neutralino, 
leads to a collider signature of OSSF leptons plus MET.
The case of stau production will be considered separately.
In the RSIII the flavor structure for the final leptons is in general different.
In the special e-philic or mu-philic cases, where the lightest \ZZ-odd fermions decay exclusively to electrons or muons, respectively,
we can extrapolate the observed exclusion limit by ATLAS for left-handed slepton pair-production~\cite{ATLAS-CONF-2013-049}
assuming that the detection efficiency of the most sensitive SR remains constant up to higher mass scales.
Taking into account the larger production cross-section for the fermions
one can estimate the lower mass exclusion limit $m_{\Sigma^\pm_1} > 630\gev$.
{In chargino pair-production, each chargino decays to a lepton and a slepton, 
which decays further to a secondary lepton and a neutralino. 
This process may lead to leptons of different flavor but
the final state has two additional neutrinos and in general softer leptons,
depending on the chosen intermediate slepton masses. 
Experimental signatures of dileptons plus MET are also obtained in chargino-neutralino production
decaying further via sleptons,
$pp \to\tilde{\chi}_1^\pm  \tilde{\chi}_2^0 \to \ell^\pm \ell^{\prime +}   \ell^{\prime -}  \tilde{\chi}_1^0  \tilde{\chi}_1^0   $,
when one of the final leptons is not detected. 
In this case both same flavor and opposite flavor leptons are expected\cite{ATLAS-CONF-2013-049}.

Among the several high energy physics tools have been developed which allow to reinterpret the 
results from the experimental collaborations at the LHC
we have chosen the package {\CheckMATE}~\cite{Drees:2013wra,Cacciari:2005hq,Cacciari:2008gp},
which allows to obtain exclusion limits
on simplified models of NP based on an increasing number of ATLAS and CMS analyses.
This package applies to the events generated by the user
the same selection cuts as in each of the included analyses by the experimental collaborations
using the fast detector simulator {\tt DELPHES}~\cite{deFavereau:2013fsa}.
Subsequently, making use of the ${\rm CL}_s$ prescription~\cite{Read:2000ru,Read:2002hq} on the most sensitive SR,
it establishes whether a given point under evaluation is ruled out or not
 based on the data given by the collaborations in their published analyses.
The implementation of the model in HEP tools is described in more detail in Sec.~\ref{sec:modelimplementation}.
The most accurate exclusion results are expected for processes with the same production and decay topologies, 
as well as similar production cross-sections,
as those in the supersymmetric searches reported in the included experimental analyses.
Notice that the cuts in the experimental analyses have been optimized for the mass range where the exclusion limits are found.

If more than one NP scalar is lighter than the produced fermions, 
 additional decay channels open up, Eqs.~(\ref{eq:sigmacascades})-(\ref{eq:sigmacascades3}),
for which there is no analogous supersymmetric process with similar decay topologies.
The heavier scalars decay further, dominantly to a gauge boson and the DM candidate. 
This secondary decay leads to large hadronic activity and is not expected to improve the exclusion sensitivity in any of the  processes included in {\CheckMATE}.
Most of the events with the additional topologies should not pass the selection cuts of the LHC analyses, 
which are optimized to reject additional hadronic activity.
Therefore, the number of selected events should decrease as the branching ratios of the new decay channels increase.
It is then natural to define a ``best case scenario'', where the \ZZ-odd fermions are the NLOP and all other NP particles are heavier, 
and a ``worst case scenario'', where all NP scalars are light and nearly degenerate.
In the latter case the branching ratio of $\Sigma_1^+$ to the heavier scalars approaches $75\%$.
It should be noticed, however, that a minimal mass splitting with the DM candidate is necessary 
in order to avoid a very large contribution of the co-annihilation channel in the early Universe.

In the intermediate case, in which the decay $A^{0}$ and $H^{\pm}$ are kinematically open but
significantly heavier than $H^{0}$, 
 the decay to the DM candidate will be enhanced with respect to the other channels.
Since the mass splitting of the two heavier scalars is strongly bounded by EWPO 
the above mentioned cases cover most of the allowed parameter space.

Within each of the benchmark scenarios discussed, the decay to leptons of the first two families has the highest sensitivity.
The case when the \ZZ-odd fermions decay predominantly to taus, which have small branching ratios to leptons, is not expected to lead to
a significant exclusion in our analysis with {\CheckMATE}, for which no experimental analyses have yet been included in this package.
This case will be considered separately, reinterpreting the stau search analysis reported in Ref.~\cite{Aad:2015eda}.

In more realistic scenarios, f.i. in cases where the decay process involves several final state topologies, 
 only the SR with the largest expected sensitivity is considered.
It is possible, however, to combine those SRs and improve the exclusion limits using the ${CL}_{s}$ method~\cite{Read:2000ru,Read:2002hq}. 

\subsection{Combination strategy}
\label{sec:combination}

In each of the decay channels, defined by their experimental signature of hard $e^{+} e^{-}$,  $\mu^{+} \mu^{-}$, and $e^{\pm} \mu^{\mp}$ plus MET,
we use the package {\CheckMATE}~\cite{Drees:2013wra,Cacciari:2005hq,Cacciari:2008gp}
to identify the most sensitive SR.
Since the flavor of the leptons depends on the unknown Yukawa couplings a realistic analysis should allow for its whole range.
In the range of masses we are considering this SR turns out to be
 $\rm{SR}$--${\rm{m}_{\rm{T}2,110}}$ of Ref.~\cite{ATLAS-CONF-2013-049},
except for $m_{\Sigma^{\pm}}\approx 350\gev$, 
where $\rm{SR}$--${\rm{m}_{\rm{T}2,110}}$ and $\rm{SR}$--${\rm{m}_{\rm{T}2,90}}$ have similar sensitivities.
We have chosen to use only the former SR. The eventual small loss in sensitivity can be regarded as conservative.

Assuming that the three dileptonic channels are uncorrelated, and thus statistically independent,
we combine these channels using the ${\it CL}_{s}$ method~\cite{Read:2002hq,Read:2000ru}, 
taking into account for the uncertainty on the background as in~Ref.\cite{Junk:1999kv}.
Details about our implementation of the $\it CL_{s}$ method are given in Appendix~\ref{sec:appCLs}.
We neglect the uncertainty on the signal since it is much smaller and therefore its effect should be subleading.
The uncertainty due to the statistics of the Monte Carlo simulations has been ignored, 
as it can be eventually reduced with larger samples~\cite{Aguilar-Saavedra:2013twa}.
The combination is expected to lead to stronger exclusion limits whenever more than one channel contributes to the final dileptons.
It should be noticed that we cannot combine the decay channels with decays to taus.

\subsection{Implementation of the model in Heptools}
\label{sec:modelimplementation}
The model has been implemented in the {\tt Mathematica} package {\tt FeynRules (v2.0)}~\cite{Alloul:2013bka}
where the derivation of the complete set of Feynman rules from the Lagrangian
given in Eq.~(\ref{eq:LRSIII}) are performed. 
The model files obtained from  {\tt FeynRules}
are exported to {\tt micrOMEGAs (v4.1.8)}~\cite{Belanger:2013oya} 
where DM observables are  evaluated. 
The model is then exported in the Universal FeynRules
Output (UFO)  format 
to the parton-level Monte Carlo (MC) generator
{\tt MadGraph (v5.2.2.3)}~\cite{Alwall:2014hca}. 
The signal events are generated at $\sqrt{s}=8 \tev$, without cuts in the run cards, 
where a total of 30K of events per point in the parameter space is simulated. 
The MC samples incorporate the NNLO~\cite{Ball:2012cx} parton distribution functions (PDF). 
{\tt MadGraph} is interfaced with {\tt Phythia (v6.4)}~\cite{Sjostrand:2006za}, 
which simulates the parton showering and hadronization.
In order to evaluate the production cross-section $pp \to \Sigma^{+}_i \Sigma^{-}_i, i=1,2,3$,
we compute  the
chargino pair-production in the pure gaugino limit with a modified version of
{\tt prospino}~\cite{Beenakker:1996ed},
 at next to the leading order (NLO) in $\alpha_s$,
where we have set to zero the chargino-quark-squark couplings in order to eliminate the t-channel contribution. 
Finally the signal samples and their corresponding NLO cross-sections 
are passed to {\tt CheckMATE (v1.1.15)}~\cite{Drees:2013wra,Cacciari:2005hq,Cacciari:2008gp}, 
where the samples pass thought a fast detector simulator {\tt DELPHES (v3.0)}\cite{deFavereau:2013fsa},
 which uses FastJet~\cite{Cacciari:2011ma} with the anti-kT algorithm~\cite{Cacciari:2008gp} 
for particle reconstruction. 

\section{Numerical results}
\label{sec:results} 

We define two benchmark scenarios which satisfy all constraints discussed in Sec.~\ref{sec:Model_Constraints},
the ``best case scenario'' ($\mathcal{S_B}$), with decoupled heavier scalars,
and the ``worst case scenario'' ($\mathcal{S_W}$), with nearly degenerate scalars,
\begin{align}
\SB &:\quad
\mdm=70{\gev}, \ \mhp=700{\gev},\ \mha=700{\gev}~,
\nonumber\\
\label{eq:SBW}
\SW &:\quad
\mdm=60.2{\gev}, \ \mhp=70.4{\gev},\ \ \, \mha=110.0{\gev}~.
\end{align}
The DM relic density lies within the measured range by Planck~\cite{Ade:2015xua},
 $\Omega_{DM}h^2=0.1197\pm 0.0044$.
The mass of the lightest \ZZ-odd charged fermion varies between its LEP lower limit, Eq.~(\ref{eq:LEPmA}), and $700\gev$.
The two heavier \ZZ-odd fermion triplets, which are not phenomenologically relevant, are set to $1.5\tev$ and $2.5\tev$, respectively.
Since the Yukawa couplings of the \ZZ-odd fields
are related to the underlying mechanism of neutrino mass generation,
a realistic phenomenological analysis of the RSIII should also study the flavor structure of the model.
We define the following extreme
 cases 
for the normalized Yukawa couplings to the lightest \ZZ-odd fermions:
{e-phobic} ($\hat Y_{1}=0$),
{mu-phobic} ($\hat Y_{2}=0$),
{e-mu-symmetric} ($\hat Y_{1}=\hat Y_{2}\le 1/\sqrt{2}$), 
and 
{ tau-philic} ($\hat Y_{1} \approx 1$),
which should be regarded as simplified models in flavor space.

\subsection{Best case scenario}
\label{sec:dileptons}
Within our benchmark scenario with decoupled heavier scalars, $\SB$, 
we have generated random parameter-sets for which the neutrino constraints are satisfied, 
and where the lightest \ZZ-odd fermion mass, $m_{\Sigma_1^\pm}$, lies within the allowed range.
The most relevant parameters are $m_{\Sigma_1^\pm}$, which determines the production cross-section at the LHC, 
and the  normalized Yukawa couplings of the triplet fermions, $\hat Y_{\alpha}$, with $\alpha=1,2,3$, 
which fully determine the tree-level branching ratios  $\mathcal{B}_\ell$, with $\ell=e,\mu\,\tau$.

The  implementation of our model in high energy physics tools has been described in Sec.~\ref{sec:modelimplementation}.
For each parameter-set we generated events for our process at $8\tev$ center of mass energy, 
$pp \to  \Sigma_{1}^+ \Sigma_{1}^-$ followed by $\Sigma_{1}^\pm \to H^0 \ell^\pm$. 
We obtain with  {\tt CheckMATE} the exclusion CL in each of the three most sensitive SRs,
 $\rm{SR}$--${\rm{m}_{\rm{T}2,110}}$ in the channels  $e^{+} e^{-}$,  $\mu^{+} \mu^{-}$,  $e^{\pm} \mu^{\mp}$ plus MET,
as well as the number of background, observed and signal events which pass all the cuts of that experimental search~\cite{ATLAS-CONF-2013-049}. 
With the latter we compute the combined exclusion confidence level with the {\it CL}$_s$ method
 described in Sec.~\ref{sec:combination}. 
For e-philic and mu-philic scenarios we have checked that both methods are consistent within the numerical uncertainties,
which in the {\it CL}$_s$ method strongly depends on the numerical integration and on the background uncertainty.

We focus on regions of parameter space for which the exclusion CL lies above $90\%$.
In Fig.~\ref{fig:BRsum.vs.diff} we show the $95\%$~CL exclusion contours in the $\mathcal{B}_e$,$\mathcal{B}_\mu$ plane (panel a) 
and in the ${\mathcal{A}_{e\mu}}$, $(\mathcal{B}_e-\mathcal{B}_\mu)$ plane (panel b),
with ${\mathcal{A}_{e\mu}} = (\mathcal{B}_e -\mathcal{B}_\mu )/(\mathcal{B}_e+\mathcal{B}_\mu)$.
\begin{figure}[t]
\begin{centering}
\hspace{-.5cm}
\includegraphics[scale=0.6]{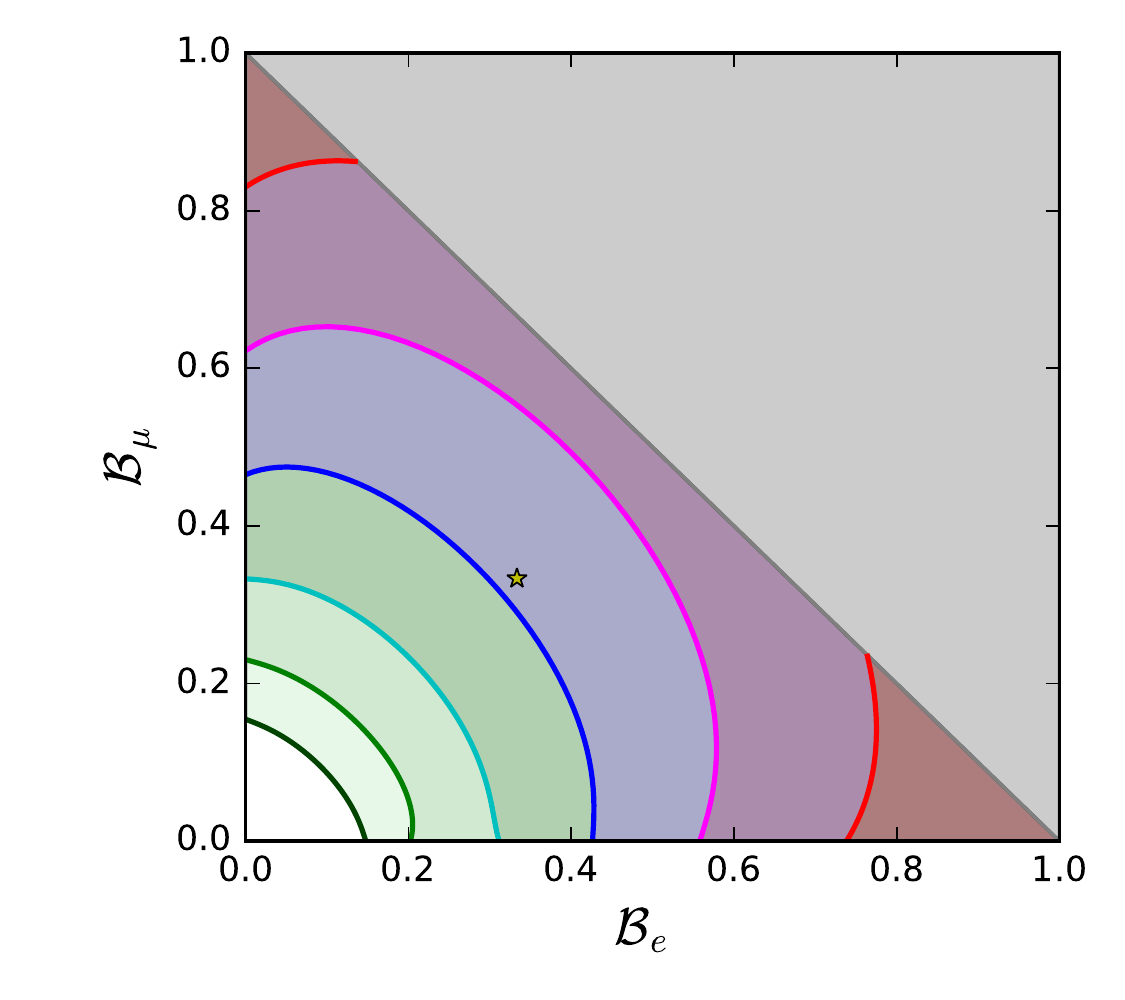}
\includegraphics[scale=0.6]{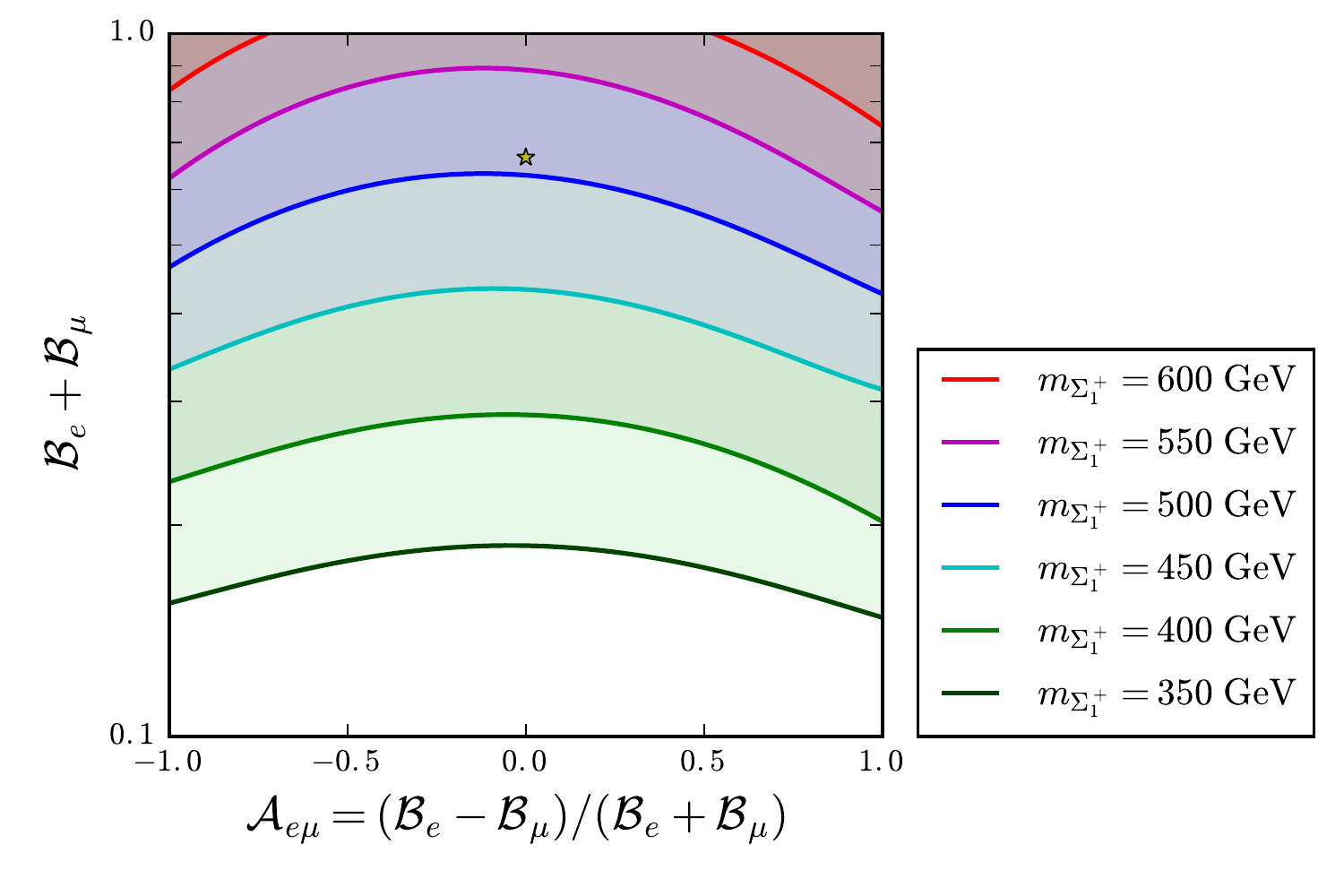}
\put(-366,180){{(a)}}
\put(-171,180){{(b)}}
\caption{\label{fig:BRsum.vs.diff} 
Contours of constant  $m_{\Sigma^+_{1}}$ 
for the present LHC exclusion sensitivity of the RSIII
in the $\mathcal{B}_{e} ,\mathcal{B}_{\mu}$ plane~(a)
and   $\mathcal{A}_{e\mu},(\mathcal{B}_{e} +\mathcal{B}_{\mu})$ plane~(b),
for $\mdm=70\gev$ 
and $\mhp\approx\mha>\msig{1}{\pm}$.
The flavor symmetric scenario with  $\mathcal{B}_{e} =\mathcal{B}_{\mu}=\mathcal{B}_{\tau}$ is shown with a 
star. 
The shaded triangle in (a) is not physical.
Both figures show the same results.
In (b) the area above each contour is excluded for the corresponding NP fermion mass.
}
\end{centering}
\end{figure}
The contours in the $95\%$~exclusion CL have been obtained fitting 
${\mathcal{A}_{e\mu}}$ as a function  of $(\mathcal{B}_{e} +\mathcal{B}_{\mu})$
with a quartic polynomial.
 The regions above the corresponding curves are excluded.
Changing the order of the fitted polynomial we conclude that
the uncertainty in these fits turns out to be larger for ${\mathcal{A}_{e\mu}}=\pm 1$.
As expected, for a given fermion mass the strongest exclusion is obtained for the mu-phobic case,
with ${\mathcal{A}_{e\mu}}=1$, 
followed by  the mu-philic case,
with  ${\mathcal{A}_{e\mu}}=-1$.
In the $e,\mu$ symmetric case, with  ${\mathcal{A}_{e\mu}}=0$ and ${\mathcal{B}_e}={\mathcal{B}_\mu}$, 
the exclusion sensitivity is reduced 
since only half of the events without taus lead to OSSF leptons, which fall into the most sensitive SRs,
while the other half of those events lead to OSDF leptons.
Shown as a 
star is the flavor symmetric case, in which all three branching ratios are equal.
As the branching ratios to taus increase, the exclusion sensitivity decreases,
since most of these events are lost in the analysis, resulting in a smaller fermion mass exclusion.
F.i., for  ${\mathcal{B}_\tau}  = 1-{\mathcal{B}_e}-{\mathcal{B}_\mu} \approx 0.85$, 
ATLAS~\cite{ATLAS-CONF-2013-049} excludes $m_{\Sigma_1^{\pm}} \lsim 350\gev$,
corresponding to the $m_{\Sigma^{\pm}}= 350\gev$ contour on the lower part of Fig.~\ref{fig:BRsum.vs.diff}b.
It should be noted that these results alone 
do not constitute solid lower mass limits for the fermions (as a function of their Yukawa couplings)
since the experimental analysis does not cover the region with compressed spectra.
We target the parameter region with small \ZZ-odd fermion masses at the end of this section 
reinterpreting a search for electroweak supersymmetric searches in the regions of compressed spectra. 
For consistency we have checked the exclusion limits obtained with {\tt CheckMATE} for small
\ZZ-odd fermion-scalar mass splitting,
where most decay leptons fail to have sufficient $p_T$ to pass the experimental cuts.
Here we set $\mdm=70\gev$ as in $\SB$.
In the most sensitive e-philic case we can exclude $m_{\Sigma_{1}}>135\gev$, i.e.\ with a mass splitting larger than $65\gev$, 
while for $\mathcal{B}_{\tau}=0.85,\ \mathcal{B}_{e}=0.15$ this mass limit increases to $m_{\Sigma_{1}}>155\gev$. 
Similar results are obtained for the e-phobic case.

\begin{figure}[h!]
\begin{centering}
\includegraphics[scale=0.625]{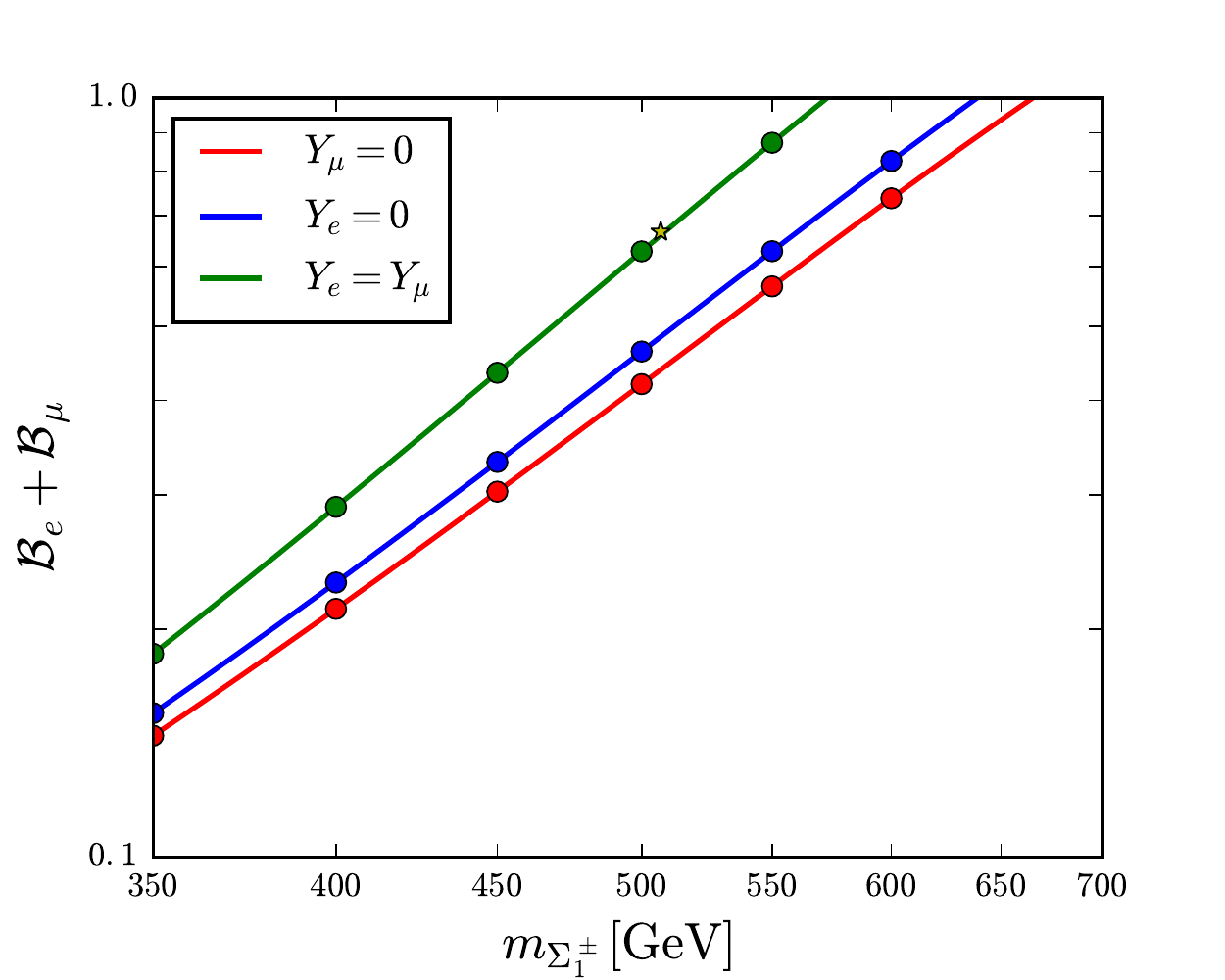}
\caption{\label{fig:BRsum.vs.mSigma} 
Present LHC exclusion sensitivity
in the $   m_{\Sigma^+_{1}}, (\mathcal{B}_{e} +\mathcal{B}_{\mu})$ plane
for $\mdm=70\gev$ and $\mhp\approx\mha>\msig{1}{\pm}$,
in the mu-phobic (red), e-phobic (blue) and e-mu-symmetric (green) scenarios.
The flavor symmetric scenario with  $\mathcal{B}_{e} =\mathcal{B}_{\mu}=\mathcal{B}_{\tau}$ is shown with a 
star. 
The region above each curve is excluded. 
}
\end{centering}
\end{figure}

The results obtained from Fig.~\ref{fig:BRsum.vs.diff} for ${\mathcal{A}_{e\mu}}=-1$, 
${\mathcal{A}_{e\mu}}=0$, 
and ${\mathcal{A}_{e\mu}}=1$ 
are shown in Fig.~\ref{fig:BRsum.vs.mSigma}, where ${\mathcal{B}_e}-{\mathcal{B}_\mu}$ is plotted as a function of $m_{\Sigma_1^{\pm}}$.
One observes that in the e-mu-symmetric case, corresponding to ${\mathcal{A}_{e\mu}}=0$, 
the mass limit is reduced by up to $50\gev$ for large masses, down to approximately $20\gev$ for the smaller masses. 
In the mu-phobic case we obtain the highest exclusion sensitivity, 
excluding masses of $\Sigma_{1}$ of up to approximately $660\gev$.

Recently ATLAS has performed a dedicated analysis~\cite{Aad:2015eda} to target compressed spectra, as well as decays with final tau leptons.
The bounds on sleptons can be reinterpreted in our model in the e-philic, mu-philic and tau-philic limits 
taking the larger production cross-sections of the \ZZ-odd fermions into account, 
since both the stau decay, $\tilde{\tau}_{1} \to \tau^{-} {\tilde{\chi}^{0}_{1}}$
and $\Sigma_{1}^{-} \to \tau^{-} H^{0}$, lead to the same experimental signature. 
In the DM region relevant for our study, 
with $m_{\tilde{\chi}^{0}_{1}}$ between $50$ and $70\gev$,
the bounds on direct stau production are not yet strong enough to reach the exclusion level.
However, rescaling the cross-section, one can safely exclude $m_{\Sigma_{1}^\pm}$ between the LEP bound of $103.5\gev$ and $300\gev$,
as shown in  Fig.~\ref{fig:XStau.vs.mSigma} for $\mdm=60\gev$ and  $\mdm=80\gev$. 
For smaller $\mdm$ these limits are stronger, allowing to extrapolate our results to the whole scalar mass range.
Assuming that the excluded cross-section for $m_{\Sigma_{1}^\pm}=300\gev$ can be extrapolated to higher masses,
implying that the sensitivity of this analysis remains constant,
this limit can be extended to exclude fermion masses below approximately $400\gev$.
\begin{figure}[t]
\begin{centering}
\includegraphics[scale=0.625]{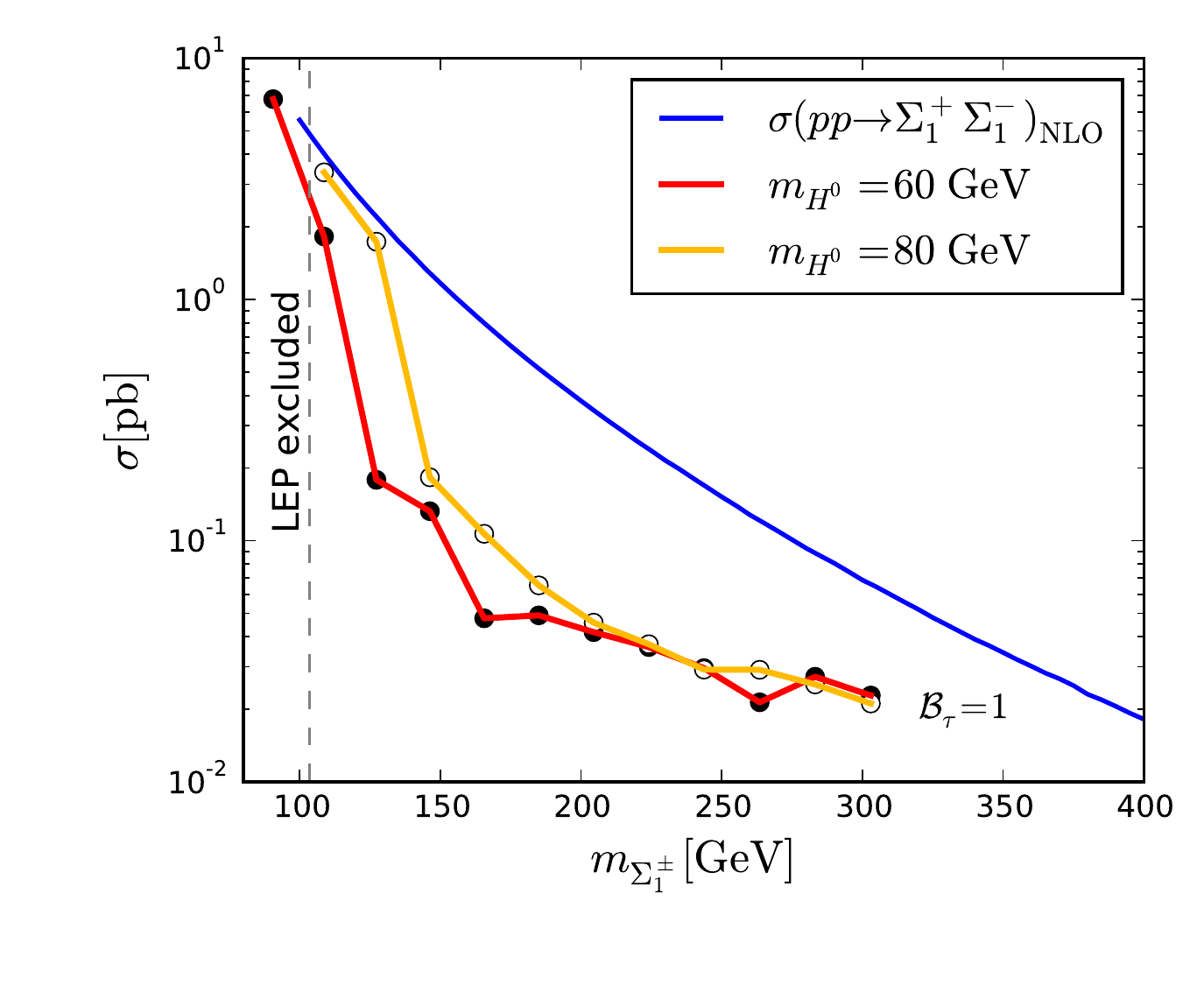}
\caption{\label{fig:XStau.vs.mSigma} 
NLO production cross-section for charged \ZZ-odd fermion pairs
 at the LHC with $8\tev$ center of mass energy (blue line) as a function of the fermion mass.
The corresponding 
$95\%$~CL exclusion limits 
for  the tau-philic case, 
when they decay exclusively to a tau and the DM scalar,
are shown for  $\mdm=60\gev$ (red line, black dots) and $80\gev$ (yellow line, white dots).
The limits have been obtained from
those derived in~\cite{Aad:2015eda} for stau pair-production.
Also shown is the LEP lower bound on $m_{\Sigma_{1}^{\pm}}$.}
\end{centering}
\end{figure}

For sleptons of the first two generations the slepton exclusion sensitivity is significantly stronger, 
allowing to exclude significant regions of parameter space~\cite{Aad:2015eda}.
Therefore, we can safely extend the limits obtained for the tau-philic case to 
the most general flavor structure.
We conclude that
all light \ZZ-odd fermion masses not covered by our previous analysis with {\tt CheckMATE} can be excluded, 
so that the exclusion limits obtained in Fig.~\ref{fig:BRsum.vs.mSigma} are solid lower mass exclusion limits
for our simplified model scenario.
\bigskip

\subsection{Worst case scenario}
\label{sec:worstcase}
The ``worst case scenario'' ($\mathcal{S_W}$), Eq.~(\ref{eq:SBW}),
has been chosen such that the heavier scalars are lighter than the produced \ZZ-odd fermions, opening additional production and decay channels at the LHC.
For a sufficiently large mass splitting between the fermion triplet and the scalars 
the branching ratios to the two neutral scalars approach $25\%$, and that of the charged scalar, the remaining $50\%$.
For instance, for $m_{\Sigma_1^\pm}=350\gev$ one obtains
\begin{align}
\sum_{\ell=e,\mu,\tau} {\rm Br}(\Sigma_1^\pm\to \ell^\pm H^0\ |\ A^0\ |\ H^\pm) =0.253\ |\ 0.234\ |\ 0.512~, 
\end{align}
i.e.\ very close to the asymptotic values.

Adding to the previously considered decay chain~(\ref{eq:sig2l})
the new decay chains of the \ZZ-odd fermions, Eqs.~(\ref{eq:sigmacascades},\ref{eq:sigmacascades2}), could in principle 
lead to new significant experimental signatures. 
In our analysis with {\CheckMATE}, however, those channels also lead to additional hadronic activity in the final state.
We observed that the experimental cuts are effective in excluding most of these events,
resulting in only a small number of new signal events from those channels.
The overall effect on the exclusion CL is small, increasing the excluded mass by less of $20\gev$,
 while the computational effort turns out to be very large.
Therefore we have neglected the new decay channels, resulting in a slightly smaller exclusion sensitivity,
and only consider the decay to the DM candidate as in the ``best case scenario''.

We focus here on 
the e-philic and mu-philic cases of scenario $\SW$, where the exclusion CL can be obtained directly from {\CheckMATE}.
In Fig.~\ref{fig:WCS.CL.vs.mSigma} we show the exclusion CL obtained with {\tt CheckMATE} 
varying  $m_{\Sigma_1^\pm}$ between $340$ and $400\gev$.
\begin{figure}[t]
\begin{centering}
\includegraphics[scale=0.7]{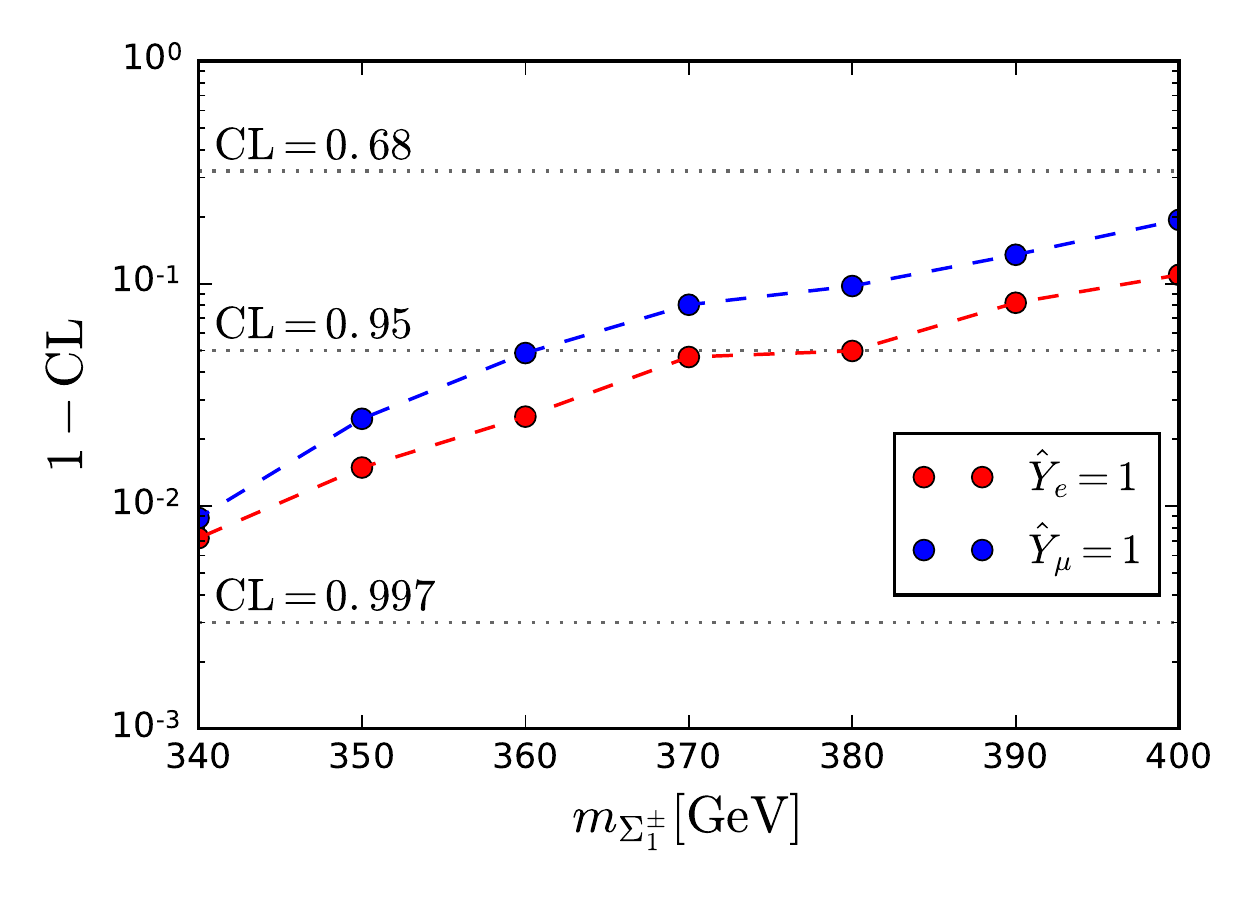}
\caption{\label{fig:WCS.CL.vs.mSigma} 
Exclusion confidence level $\rm{CL}$ as a function of the NP fermion mass
in the ``worst case scenario'' of {Eq.~(\ref{eq:SBW})}. 
The dots correspond to 
the e-philic case $\hat Y_e=1$ (red),
mu-philic case  $\hat Y_\mu=1$ (blue).
The dashed lines simply connect the dots.
Masses for which $1-\rm{CL}<0.05$ are excluded.}
\end{centering}
\end{figure}
Only one scenario for each fermion mass has been computed here. 
We observe that, retaining only around $25\%$ of the events, the masses of between
$360\gev$ for the mu-philic case, and $380\gev$ for the e-philic case.

\section{\label{sec:level5} Summary and Conclusions}

We have explored the Radiative Type III Seesaw model (RSIII), 
a scotogenic model in which an additional scalar doublet and at least two fermion triplets 
of $SU(2)_L$, odd under a conserved \ZZ\ global symmetry, 
are added to the SM.
This model has a natural DM candidate, the LOP, and radiatively generates the neutrino masses by an effective Weinberg operator.
We have focused in the low mass scalar DM region, where the  LOP is a viable DM candidate
satisfying all present theoretical and experimental constraints.
In this region of parameter space
the \ZZ-odd fermion triplets can have masses above the LEP limit for wino-like charginos, 
 potentially leading to new physics signatures at the LHC. 
In order to set solid exclusion limits on the model we identify two extreme scenarios, 
a ``best case scenario'' where only the DM candidate is lighter than the fermion triplet,
and a ``worst case scenario'' where all scalars are light.
In the former, the decay process has simple decay topologies, 
which have been already studied in simplified model spectra analyses of supersymmetric searches at the LHC.
In the latter, new decay channels open up, leading to longer decay chains and more complex experimental signatures.
These two benchmark scenarios can be regarded as limiting cases,
with ``intermediate scenarios'', where the heavier scalar masses lie 
in-between those values,
leading to exclusion limits which lie within the two extreme cases.
For these scenarios we have analyzed the present theoretical and experimental constraints.

We reinterpret a set of experimental searches for supersymmetric particles at the LHC by 
ATLAS
~\cite{ATLAS-CONF-2013-049,Aad:2014nua,Aad:2014vma,
Aad:2015eda}
within the framework of the RSIII with help of the package {\tt CheckMATE}~\cite{Drees:2013wra,Cacciari:2005hq,Cacciari:2008gp}.
In order to do this we implemented the model in high energy physics tools and generated the NP events 
which are then processed further by {\tt CheckMATE}.
The process with the most sensitive signature turns out to be pair-production of charged NP fermions, 
 decaying each to the DM candidate and an electron or a muon.
The resulting experimental signature, opposite sign dileptons plus MET,
 is also obtained in two supersymmetric processes: 
  slepton pair-production decaying to the LSP and a lepton,
or chargino-neutralino pair-production decaying subsequently via intermediate sleptons, where one of the charged leptons is lost in the detector.
The fermion triplets decay via Yukawa couplings to a lepton and a scalar. 
Since these Yukawa couplings are intrinsically related to the neutrino mass matrix,
a determination of the flavor structure of the final state would allow to 
directly study neutrino properties at colliders.
It is therefore highly relevant to obtain exclusion limits as a function of the flavor structure of the final state.
We have expressed those limits as a function of the branching ratio of the charged \ZZ-odd fermion 
to the DM candidate plus an electron or a muon.
In the ``best case scenario'', with decoupled heavy scalars,
the strongest limits on the \ZZ-odd fermion triplets are obtained in the e-philic case, for which we exclude masses below roughly $660\gev$.
This limit is reduced to 
$640\gev$ and $570\gev$, in, respectively,  
 the mu-philic and e-mu symmetric cases.
One should notice that our results are subject to uncertainties of the Monte Carlo simulations of the analysis which may be reduced with higher statistics. 
For light NP fermions, below roughly $150\gev$, the dilepton searches included in {\tt CheckMATE} fail to exclude our model.
In order to obtain solid lower limits on the \ZZ-odd fermion masses we recast an analysis by ATLAS~\cite{Aad:2015eda} 
for searches in the compressed mass spectra region.
The experimental results included in {\tt CheckMATE} are not sensitive to final state taus, 
which mostly generate hadron activity excluded  in their cut-based analyses.
We recast the results of~\cite{Aad:2015eda} for tau searches, taking into account the larger cross-sections for fermion pair-production, to obtain a lower mass limit of around $400\gev$ for fermion triplets in the tau-philic case.

In the ``worst case scenario'' we have obtained limits both including only the primary decays to the DM candidate, and including all channels.
The results in both cases are consistent with each other, with a slight gain in exclusion sensitivity 
in latter case, 
albeit at the price of a huge increase in computational effort. 
We have therefore restricted our analysis to the former case. The branching ratios are reduced by a factor of almost four,
reducing the sensitivity to the level of slepton searches.
In the e-philic and mu-philic cases we can exclude fermion triplet masses below roughly $380$ and $360\gev$, respectively. 
As in the ``best case scenario'', the lower mass region is excluding by a recast of the compressed spectra analysis~\cite{Aad:2015eda}.
For the tau-philic case no limits can yet be set.
\medskip

The  LHC exclusion limits obtained in flavor space on our scotogenic model, the RSIII, 
should be easily extended to all NP models with NLOP fermions in the adjoint representation of $SU(2)_L$ decaying to a scalar DM candidate and a lepton.

\subsection*{Acknowledgments}

We thank  
J.S.~Kim,
E.~Rojas,
and
J.D.~Ruiz Alvarez
for helpful discussions.  
D.R.\ and O.Z.\ have been partially supported by  COLCIENCIAS through the Grants No.\ 111-556-934918 and 111-565-84269.


\begin{appendix}

\section{CL$_s$ method}
\label{sec:appCLs}
In order to obtain 
the exclusion limit for ${N_{}} $ experimental channels 
we combine them with the $\it CL_{s}$ method defined in Ref.~\cite{Read:2000ru,Read:2002hq}.
We take into account the leading uncertainty 
from the background
convoluting the individual channel likelihoods
${\mathcal L}(n_k; s_k+b_k) $ and ${\mathcal L}(n_k; b_k) $
for the signal plus background and background hypotheses, 
respectively, 
 with a Gaussian distribution with standard deviation $\sigma_{ b_k}$,
\begin{align} 
\label{eq:avgL}
\langle {\mathcal L}(n_k; s_k+b_k) \rangle &= 
\frac{1}{\sqrt{2 \pi  }  \sigma_{ b_k} }
\int_0^\infty d b_k^\prime \exp\left(- \dfrac{(b_k^\prime-b_k)^2}{2 \sigma_{ b_k}^2}\right)
\dfrac{ e^{-(s_k+b_k^\prime)}(s_k+b_k^\prime)^{n_k} }{n_k!}~,
\end{align} 
with $\langle {\mathcal L}(n_k; b_k) \rangle $ defined analogously.
Here $n_k$, $s_k$ and $b_k$ denote, respectively,
the number of events, the expected  signal events, and the corresponding background events in each channel. 

The likelihood ratio test-statistics function is given by
\begin{eqnarray}{{\label{eq:test_statistics}}}
 && Q = 
\prod_{i=1}^{N_{}} 
\left( \dfrac{ e^{-(s_i+b_i)}(s_i+b_i)^{n_i} /n_i! }{ e^{-b_i}b_i^{n_i} /n_i! } \right) 
= e^{- s_{\rm tot}} \prod_{i=1}^{N_{}} 
\left( 1+ \dfrac{ s_i }{ b_i } \right)^{n_i}~,
\end{eqnarray}
with $s_{\rm tot}=\sum_k^{N_{n}} s_k$.
The observed likelihood ratio test statistics $Q_{\rm obs}$ is defined
 analogously setting $n_i=n^{\rm obs}_i$, 
the observed number of events reported in the experimental analyses.
The test statistics function $Q$ should also be averaged by the Gaussian distribution.
To simplify the numerical evaluation we average $\log Q$ as in  Eq.~(\ref{eq:avgL})
\begin{eqnarray}{{\label{eq:avg_Q}}}
 && \bar Q  \equiv 
\exp (\langle \log Q \rangle  )~.
\end{eqnarray}

The confidence level for exclusion ${\rm CL}=1-CL_{s}$ is given by
\begin{equation} 
CL_{s} =
\frac{CL_{s+b} }{CL_{b}}~,
\end{equation} 
with
\begin{align} 
CL_{s+b} &= 
\sum_{\bar Q<\bar Q_{\rm obs}  } \prod_{k=1}^{N_{}} 
\langle 
{\mathcal L}(n_k; s_k+b_k) 
\rangle~,
\\
CL_{b} &= 
\sum_{\bar Q<\bar Q_{\rm obs}} \prod_{k=1}^{N_{}} 
\langle 
{\mathcal L}(n_k; b_k) 
\rangle~.
\end{align} 

\end{appendix}
\medskip

\bibliography{} 
\bibliographystyle{h-physrev4}

\end{document}